\documentclass[runningheads]{llncs}

\usepackage{amsmath} 
\usepackage{amssymb} 
\usepackage{amsfonts}
\usepackage[dvipsnames]{xcolor}
\usepackage{graphicx} 
\usepackage{tabularx}
\usepackage{blkarray}
\usepackage{physics} 
\usepackage{caption}
\usepackage{subcaption}
\usepackage{stackrel}
\usepackage{mathtools}
\usepackage{mathrsfs}
\usepackage{multirow}
\usepackage[ruled,vlined,linesnumbered,resetcount]{algorithm2e}
\DontPrintSemicolon
\usepackage{etoolbox}
\makeatletter
\patchcmd{\@algocf@start}
  {-1.5em}
  {0pt}
  {}{}
\makeatother
\usepackage{booktabs}
\usepackage{bbm}
\usepackage[T1]{fontenc}

\usepackage{thm-restate}

\makeatletter
\RequirePackage[bookmarks,unicode,colorlinks=true,breaklinks=true]{hyperref}%
   \def\@citecolor{blue}%
   \def\@urlcolor{blue}%
   \def\@linkcolor{blue}%

\def\orcidID#1{\smash{\href{http://orcid.org/#1}{\protect\raisebox{-1.25pt}{\protect\includegraphics{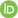}}}}}
\makeatother

\def\techreport{}

\newcommand{\sectref}[1]{Section~\ref{#1}}

\newcommand{\figref}[1]{Figure~\ref{#1}}
\newcommand{\tabref}[1]{Table~\ref{#1}}

\newcommand{\eqnref}[1]{Equation~\ref{#1}}

\newcommand{\lemref}[1]{Lemma~\ref{#1}}
\newcommand{\defref}[1]{Definition~\ref{#1}}
\newcommand{\agref}[1]{Algorithm~\ref{#1}}

\newcommand{\figfigref}[2]{Figures~\ref{#1} and \ref{#2}}

\newcommand{\startpara}[1]{{\vskip5pt\noindent{\bf #1.}}} 
\renewcommand{\url}[1]{{\def~{\char126}\sf#1}}

\newcounter{exampcount}
\newenvironment{examp}
{\refstepcounter{exampcount}
\vskip6pt\noindent
{\bf Example \arabic{exampcount}.}}
{\hfill$\blacksquare$\vskip6pt}

\DeclareMathOperator*{\argmin}{arg\,min}
\DeclareMathOperator{\opt}{opt}
\newcommand\ra{\rightarrow}

\def\deq{:\stackrel{{\tiny D}}{=}}

\def\Nset{\mathbb{N}}
\def\Nsetinf{\mathbb{N}_{\infty}}
\def\Rset{\mathbb{R}}

\def\Qset{\mathbb{Q}}

\def\cA{{\mathcal{A}}}

\def\cF{{\mathcal{F}}}
\def\cO{{\mathcal{O}}}
\def\E{\mathbb{E}}

\newcommand\Prob{{\Pr}}
\def\exp{{\mathbb{E}}}
\def\var{{\mathrm{Var}}}
\def\sd{{\mathrm{s.d.}}}
\def\mode{{\mathrm{mode}}}
\def\VaR#1{{\mathsf{VaR}_{#1}}}
\def\CVaR#1{{\mathsf{CVaR}_{#1}}}

\def\Vmax{{\mathit{V}_\mathrm{max}}}
\def\Vmin{{\mathit{V}_\mathrm{min}}}

\newcommand{\dtmc}{{\mathcal{D}}}
\newcommand{\mdp}{{\mathcal{M}}}
\newcommand{\mdpb}{{\mathcal{M}^\mathsf{b}}}
\newcommand{\dtmctuple}{{(S, \sinit, P, \AP, \lab)}}
\newcommand{\mdptuple}{{(S, \sinit, \Act, P, \AP, \lab)}}
\newcommand{\mdpbtuple}{{(S \times B, \{\sinit\}\times B, \Act, P^\mathsf{b}, \AP, \lab^\mathsf{b})}}
\newcommand{\sinit}{{s_0}}
\newcommand{\binit}{{\bar{b}}}
\newcommand{\lab}{{L}}
\newcommand{\Act}{{A}}
\newcommand{\rew}{r}
\newcommand\strat{{\pi}}
\newcommand\strats{{\Sigma}}
\newcommand\policy{{\strat}}
\newcommand\policies{{\strats}}
\newcommand{\last}{\mathit{last}}
\newcommand{\ipaths}{\mathit{IPaths}}
\newcommand{\fpaths}{\mathit{FPaths}}
\newcommand{\ipat}{\omega}
\newcommand{\fpat}{\omega}
\newcommand\saug[2]{\langle {#1},{#2} \rangle}

\newcommand\AP{{\mathit{AP}}}

\newcommand\sat{{\,\models\,}}

\def\future{{\mathtt{F}\ }}
\def\futureop{{\mathtt{F}}}

\newcommand\rewopR{{\mathtt R}}
\newcommand\rewop[3]{\rewopR^{#1}_{\,#2}[\,{#3}\,]}

\newcommand{\rvprop}{\mathsf{f}}

\begin{document}
\title{Distributional Probabilistic Model Checking}


\author{Ingy Elsayed-Aly\inst{1}\,\orcidID{0000-0003-0623-2323} 
\and
David Parker\inst{2}\,\orcidID{0000-0003-4137-8862} 
\and
Lu Feng\inst{1}\,\orcidID{0000-0002-4651-8441}
}

\authorrunning{I. Elsayed-Aly et al.}

\institute{University of Virginia, Charlottesville VA 22904, USA \\
\email{\{ie3ne,lu.feng\}@virginia.edu} \and
University of Oxford, Oxford, UK \\
\email{david.parker@cs.ox.ac.uk}}

\maketitle


\begin{abstract}
Probabilistic model checking provides formal guarantees for stochastic models relating to a wide range of quantitative properties, such as runtime, energy consumption or cost. But this is typically with respect to the \emph{expected} value of these quantities, which can mask important aspects of the full probability distribution, such as the possibility of high-risk, low-probability events or multimodalities. We propose a \emph{distributional} extension of probabilistic model checking, for discrete-time Markov chains (DTMCs) and Markov decision processes (MDPs). We formulate distributional queries, which can reason about a variety of distributional measures, such as variance, value-at-risk or conditional value-at-risk, for the accumulation of reward or cost until a co-safe linear temporal logic formula is satisfied. For DTMCs, we propose a method to compute the full distribution to an arbitrary level of precision, based on a graph analysis and forward analysis of the model.  For MDPs, we approximate the optimal policy using distributional value iteration.  We implement our techniques and investigate their performance and scalability across a range of large benchmark models. 

\end{abstract}

\section{Introduction} \label{sec:intro} 
Computer systems are increasingly being integrated seamlessly with sensing,
control and actuation of the physical world.
Many of these systems (e.g., robotics) exhibit probabilistic and non-deterministic behavior
due to inherent uncertainty (e.g., sensor noise, human interactions),
which pose significant challenges for ensuring their safe, reliable, timely and resource-efficient execution.

\emph{Probabilistic model checking} offers a collection of techniques for modelling systems that exhibit probabilistic and non-deterministic behavior. It supports not only their verification against specifications in temporal logic, but also synthesis of optimal controllers (policies).
Commonly used models include discrete-time Markov chains (DTMCs) and Markov decision processes (MDPs).
A range of verification techniques for these, and other models, are supported
by widely used probabilistic model checkers such as PRISM~\cite{KNP11} and Storm~\cite{DJKV17}.

To capture the range of quantitative correctness specifications needed in practice,
it is common to reason about \emph{rewards} (or, conversely, \emph{costs}).
Examples include checking the worst-case execution
time of a distributed coordination algorithm,
or synthesizing a controller that guarantees the minimal
energy consumption for a robot to complete a sequence of tasks.
Typically the \emph{expected} value of these quantities is computed,
but in some situations it is necessary to consider the full probability distribution.
Notably, in safety-critical applications, it can be important to synthesize
\emph{risk-sensitive} policies, that avoid high-cost, low-probability events,
which can still arise when minimizing expected cost.
Risk-aware distributional measures such as \emph{conditional value-at-risk} ($\CVaR{}$)~\cite{majumdar2020should}
address this by minimizing the costs that occur
above a specified point in the tail of the distribution.
Within probabilistic model checking, the use of \emph{quantiles}
has been proposed~\cite{UB13,randour2017percentile,klein2018advances,HJKQ20}
to reason about cost or reward distributions.


In this paper, we develop a \emph{distributional probabilistic model checking} approach,
which computes and reasons about the full distribution
over the reward associated with a DTMC or MDP.
More precisely, we consider the reward accumulated until a 
specification in co-safe LTL is satisfied,
the latter providing an expressive means to specify, for example,
a multi-step task to be executed by a robot~\cite{kressgazit}, 
or a sequence of events leading to a system failure.
We propose a temporal logic based specification for such distributional queries.

For a DTMC, we perform model checking of these queries by
generating a precise representation of the distribution,
up to an arbitrary, pre-specified level of accuracy
(the distribution is discrete, but often has countably infinite support,
so at least some level of truncation is typically required).
This is based on a graph analysis followed by a forward numerical computation.
From this, we can precisely compute a wide range of useful properties,
such as the mean, variance, mode or various risk-based measures.

For an MDP, we instead aim to optimize such properties over all policies.
In this paper, we focus on optimizing the expected value or $\CVaR{}$,
whilst generating the full reward distribution for each state of the MDP.
This is done using \emph{distributional value iteration} (DVI)~\cite{bookdr2022},
which can be seen as a generalization of classical value iteration.
Rather than computing a single scalar value (e.g., representing the optimal expected reward)
for each MDP state,
DVI associates a full distribution with each state,
replacing the standard Bellman equation with a distributional Bellman equation.

We consider two types of DVI algorithms, namely \emph{risk-neutral} DVI for optimizing the expected value 
and \emph{risk-sensitive} DVI for optimizing $\CVaR{}$. 
Risk-neutral DVI can be shown to converge to a deterministic, memoryless optimal policy, if a unique one exists~\cite{bookdr2022}.
For $\CVaR{}$, memoryless policies do not suffice for optimality,
but risk-sensitive DVI does converge for a product MDP that incorporates a (continuous)
slack variable representing a cost/reward budget~\cite{bauerle2011markov}.
For computational tractability, we present a risk-sensitive DVI algorithm based on a discretization of the slack variable,
and show that the algorithm converges to a $\CVaR{}$ optimal policy for increasingly precise discretizations.

For both DVI algorithms, in practice it is necessary
to use approximate distributional representations.
We consider the use of categorical and quantile representations.
This can impact both optimality and the precision of computed distributions
but, for the latter, we can construct the DTMC induced by generated MDP policies
and use our precise approach to generate the correct distribution.
Finally, we implement our distributional probabilistic model checking framework
as an extension of the PRISM model checker~\cite{KNP11}
and explore the feasibility and performance of the techniques on a range of benchmarks.

\vskip6pt
\noindent
\ifthenelse{\isundefined{\techreport}}{%
An extended version of this paper, with proofs, is available as \cite{dpmc-arxiv}.
}{%
This paper is an extended version, with proofs, of \cite{dpmc-nfm}.
}

\vspace*{-0.5em}
\subsection{Related Work} \label{sec:related} 
\vspace*{-0.5em}
\startpara{Distributional properties}
Some existing probabilistic model checking methods consider distributional properties beyond expected values,
notably \emph{quantiles}~\cite{UB13,randour2017percentile,klein2018advances,HJKQ20},
i.e., optimal reward thresholds which guarantee that the maximal or minimal probability
of a reward-bounded reachability formula meets a certain threshold.
While~\cite{UB13}~and~\cite{randour2017percentile} focus on complexity results,
\cite{klein2018advances} and~\cite{HJKQ20} consider practical implementations to compute quantiles,
for single- and multi-objective variants, respectively, using model unfoldings over ``cost epochs'';
\cite{HJKQ20} also proposes the use of interval iteration to provide error bounds.
By contrast, our methods derive the full distribution,
rather than targeting quantiles specifically,
and our DTMC approach derives error bounds from a forward computation.
We also mention \cite{CJJW22}, which computes probability distributions in a forwards manner,
but for infinite-state probabilistic programs and using generating functions,
and~\cite{BCFK17}, which proposes an algorithm (but not implementation)
to compute policies that trade off expected mean payoff and variance.

\startpara{Risk-aware objectives}
For MDPs, we focus in particular on \emph{conditional value-at-risk} ($\CVaR{}$).
There are alternatives, such as mean-variance~\cite{sobel1982variance} and value-at-risk~\cite{filar1995percentile} but,
as discussed in~\cite{majumdar2020should}, these are not \emph{coherent risk metrics},
which may make them unsuitable for rational decision-making.
Other work on the $\CVaR{}$ objective includes:
\cite{kvretinsky2018conditional}, which studies decision problem complexity,
but for mean-payoff rewards and without implementations;
~\cite{chow2015risk}, which repeatedly solves piecewise-linear maximization problems,
but has limited scalability, taking over 2 hours to solve an MDP with about 3,000 states;
and \cite{meggendorfer2022risk}, which proposes both linear programming and
value iteration methods to solve $\CVaR{}$ for MDPs and DTMCs.
Other, not directly applicable, approaches tackle constrained problems
that incorporate the $\CVaR{}$ objective~\cite{rigter2022planning,borkar2014risk,chow2014algorithms}.
Again, our approach differs from all these in that it computes the full distribution,
allowing multiple distributional properties to be considered.
We also work with temporal logic specifications.
Alternative temporal logic based approaches to risk-aware control
include \cite{CT18}, which proposes risk-aware verification of MDPs using cumulative prospect theory,
and \cite{JRSS18} which proposes chance constrained temporal logic for control of deterministic dynamical systems.

\startpara{Distributional reinforcement learning}
Our work is based on probabilistic model checking, which fully explores known models,
but our use of DVI is inspired by \emph{distributional reinforcement learning}~\cite{bookdr2022},
which can be used to learn risk-sensitive policies and improve sample efficiency
(see~\cite{lyle2019comparative} for a comparison of expected and distributional methods).
We take a formal verification approach and use numerical solution, not learning,
but adopt existing categorical and quantile distributional approximations
and our risk-neutral DVI algorithm is a minimization variant adapted from~\cite{bookdr2022}.
Risk-sensitive DVI is also sketched in~\cite{bookdr2022}, based on~\cite{bauerle2011markov},
but only a theoretical analysis of the method is given,
without considering practical implementation aspects,
such as how to discretize slack variables for computational efficiency,
and how such approximations would affect the correctness of model checking.
We extend risk-sensitive DVI with a discretized slack variable and show its effects theoretically in \sectref{sec:cvar} and empirically via computational experiments in \sectref{sec:exp}.

\section{Background} \label{sec:bg} 
We begin with some background on random variables, probability distributions,
and the probabilistic models used in this paper.
We let $\Nset$, $\Rset$, and $\Qset$ denote the sets of
naturals, reals and rationals, respectively,
and write $\Nsetinf=\Nset\cup\{\infty\}$.


\subsection{Random Variables and Probability Distributions}\label{sec:dist}

Let $X: \Omega \to \Rset$ be a random variable over a probability space $(\Omega, \cF, \Pr)$. 
The \emph{cumulative distribution function} (CDF) of $X$ is denoted by $\cF_X(x) := \Pr(X \le x)$, 
and the inverse CDF is $\cF_X^{-1}(\tau) := \inf\{x \in \Rset: \cF_X(x) \ge \tau\}$.
Common properties of interest for $X$ include, e.g.,
the \emph{expected value} $\exp(X)$, the \emph{variance} $\var(X)$
which is the square of the \emph{standard deviation} (s.d.), or the \emph{mode}.

\begin{figure}[t]
\centering
\hspace{-1.2 em}
\subfloat [True distribution]{
  \includegraphics[width=.33\linewidth]{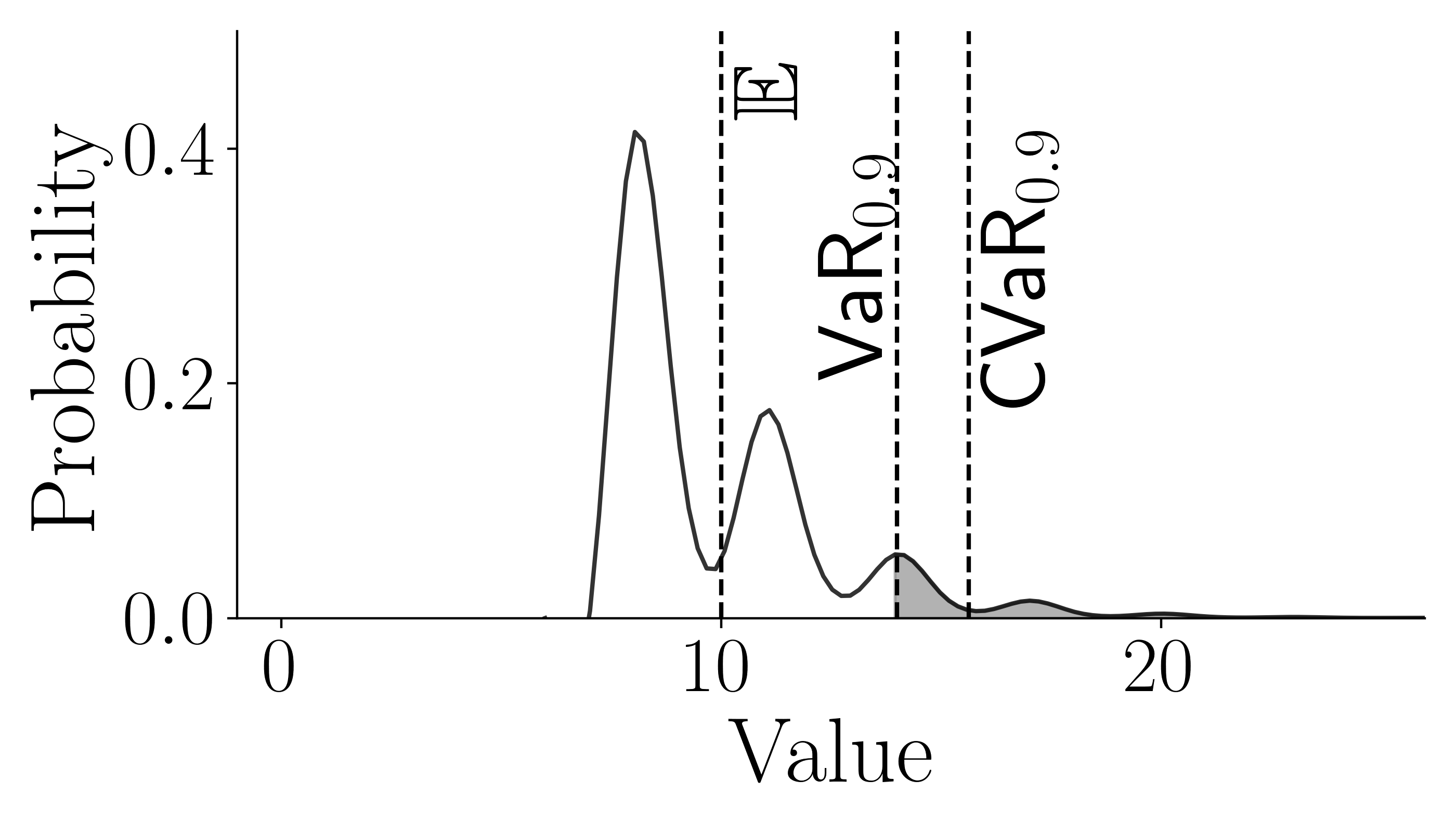}
  \label{fig:rep_dtmc_dst}
}\hspace{-1.8 em} %
\subfloat [Categorical ($m=11$)]{
  \includegraphics[width=.33\linewidth]{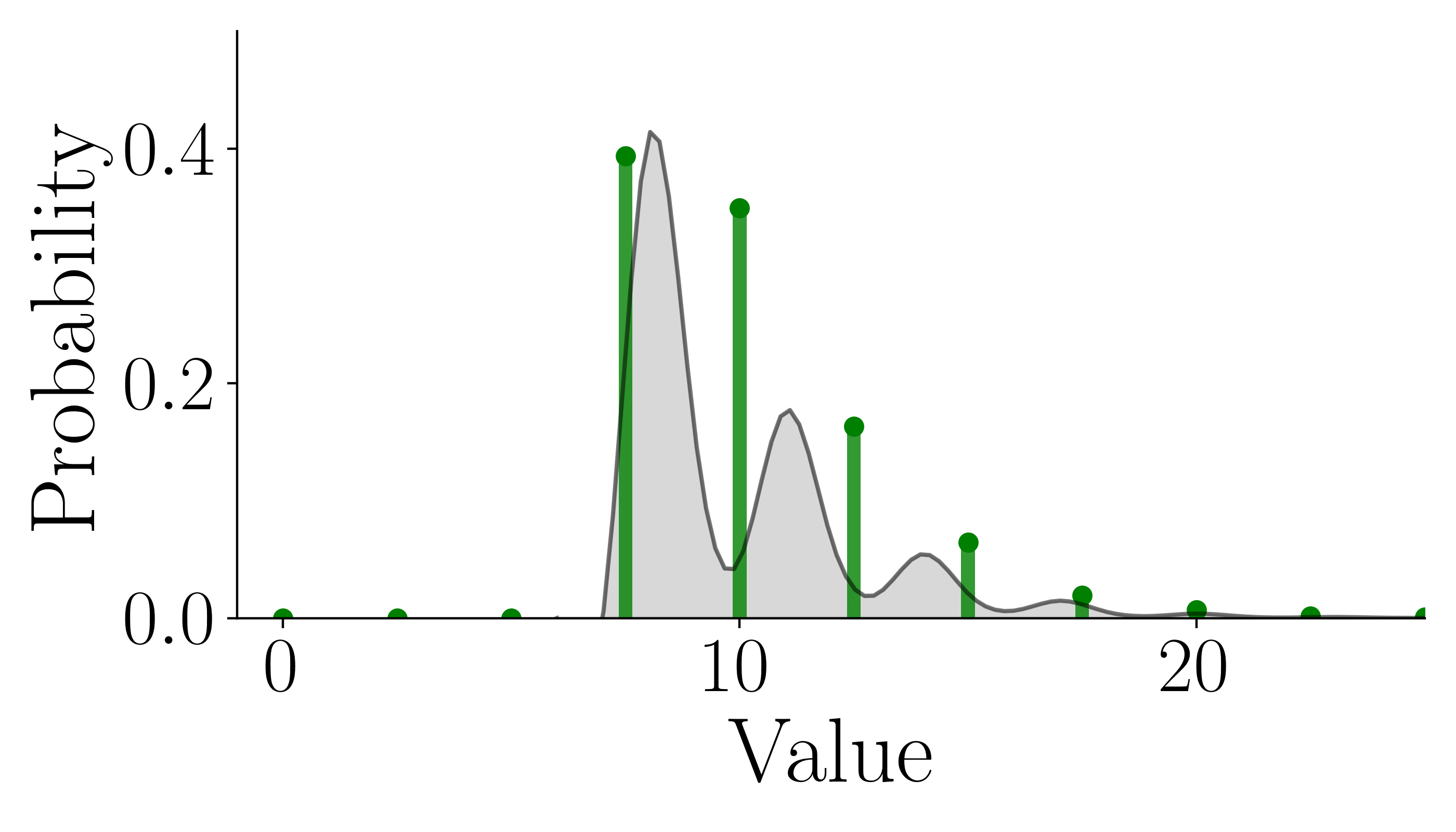}
  \label{fig:rep_c51_dst}
}\hspace{-1.8 em} %
\subfloat [Quantile ($m=10$)]{
  \includegraphics[width=.33\linewidth]{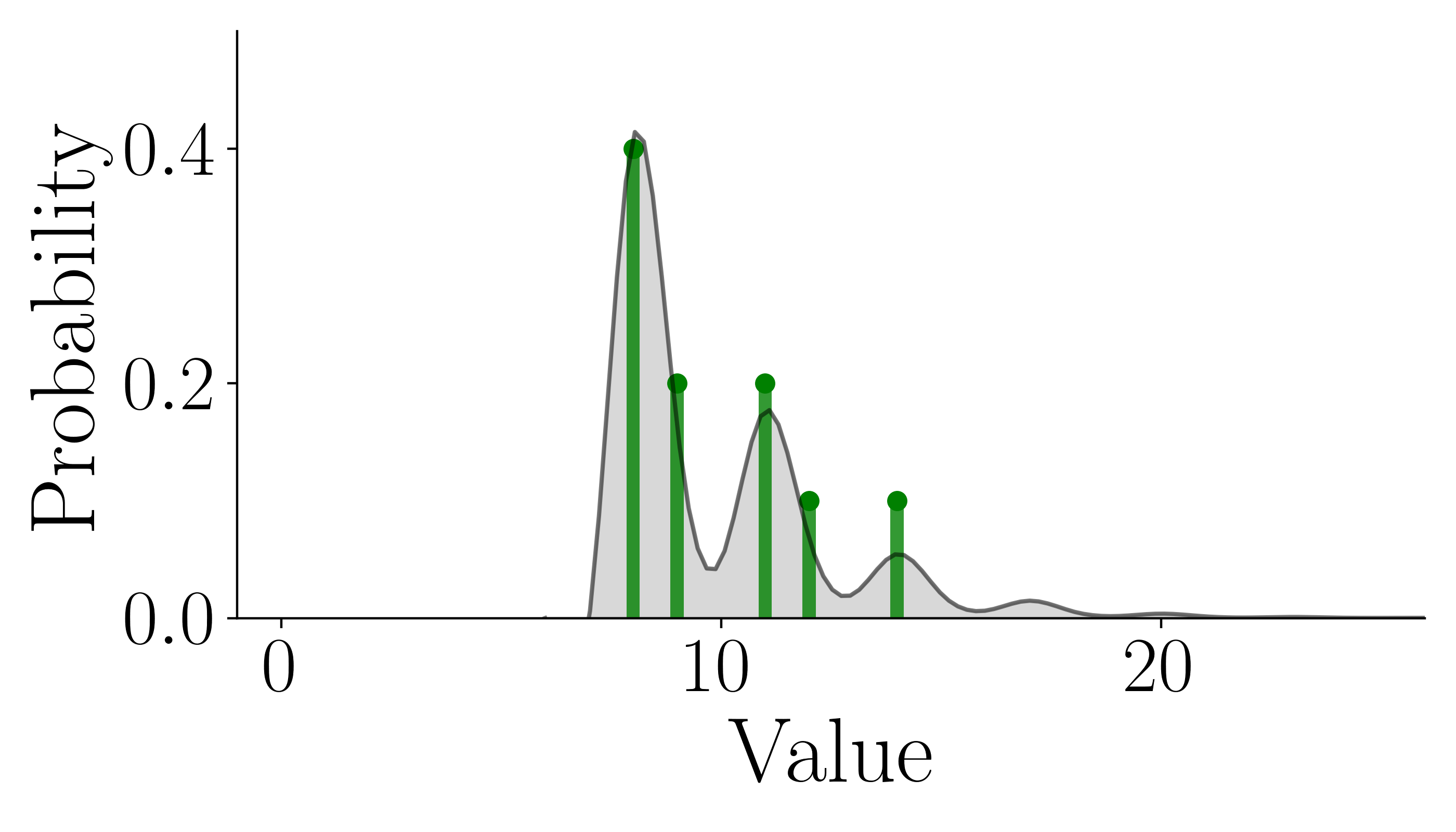}
  \label{fig:rep_qr_dst}
} %
\vspace*{-0.4em}
\caption{An example distribution with its categorical and quantile representations.}
\label{fig:dst_eg}
\end{figure}

In this paper, we also consider several \emph{risk}-related measures.
The \emph{value-at-risk} of $X$ at level $\alpha \in (0,1)$ is defined by 
$\VaR{\alpha}(X) := \cF_X^{-1}(\alpha)$, 
which measures risk as the minimum value encountered in the tail of the distribution with respect to a risk level $\alpha$. 
The \emph{conditional value-at-risk} of $X$ at level $\alpha \in (0,1)$ is given by
$\CVaR{\alpha}(X) := \frac{1}{1 - \alpha}\int_{\alpha}^1 \VaR{\nu}(X) d \nu$,
representing the expected loss given that the loss is greater or equal to $\VaR{\alpha}$.
\figref{fig:rep_dtmc_dst} illustrates an example probability distribution of a random variable $X$, annotated with its expected value $\exp(X)$, value-at-risk $\VaR{0.9}(X)$ and conditional value-at-risk $\CVaR{0.9}(X)$.

When working with the probability distributions for random variables,
we write distributional equations as $\smash{X_1 \deq X_2}$, denoting equality of probability laws
(i.e., the random variable $X_1$ is distributed according to the same law as $X_2$).
We use $\delta_\theta$ to denote the Dirac delta distribution
that assigns probability 1 to outcome $\theta \in \Rset$. 
In practice, even when distributions are discrete,
we require approximate, finite representations for them.
In this paper, we consider \emph{categorical} and \emph{quantile} distributional representations,
both of which provide desirable characteristics such as tractability and expressiveness~\cite{bookdr2022}.

\begin{definition}[Categorical representation]\label{def:cgr}
A \emph{categorical representation} parameterizes the probability of $m$ \emph{atoms} as a collection of evenly-spaced locations $\theta_1 < \cdots < \theta_m \in \Rset$.
Its distributions are of the form
$\sum_{i=1}^m  p_i \delta_{\theta_i}$ where $p_i \ge 0$ and $\sum_{i=1}^m p_i=1$.
We define the \emph{stride} between successive atoms as $\varsigma_m = \frac{\theta_m-\theta_1}{m-1}$.
\end{definition}

\begin{definition}[Quantile representation]\label{def:qtr}
A \emph{quantile representation} parameterizes the location of $m$ equally-weighted atoms.
Its distributions are of the form $\frac{1}{m} \sum_{i=1}^m \delta_{\theta_i}$ for $\theta_i \in \Rset$.
Multiple atoms may share the same value. 
\end{definition}

\figfigref{fig:rep_c51_dst}{fig:rep_qr_dst} show categorical and quantile representations, respectively, approximating the distribution shown in \figref{fig:rep_dtmc_dst}. 
%
When performing operations on distributions (e.g., during DVI),
the intermediate result might not match the chosen representation parameters. 
In that case, the result is \emph{projected} back onto the chosen representation as described in~\cite{bookdr2022}.

\subsection{Markov Chains and Markov Decision Processes}

In this paper, we work with both discrete-time Markov chains (DTMCs)
and Markov decision processes (MDPs).

\begin{definition}[DTMC]
A \emph{discrete-time Markov chain} (DTMC) is a tuple $\dtmc = \dtmctuple$, where
$S$ is a set of states, $s_0\in S$ is an initial state,
$P:S\times S\ra[0,1]$ is a probabilistic transition matrix satisfying
$\forall s\in S: \sum_{s'\in S}P(s,s')=1$,
$\AP$ is a set of atomic propositions and $\lab:S\ra2^{\AP}$ is a labelling function.
\end{definition}

A DTMC $\dtmc$ evolves between states, starting in $s_0$,
and the probability of taking a transition from $s$ to $s'$ is $P(s,s')$.
An (infinite) \emph{path} through $\dtmc$ is a sequence of states
$s_0 s_1 s_2\dots$ such that $s_i\in S$ and $P(s_i,s_{i+1})>0$ for all $i\geq 0$,
and a finite path is a prefix of an infinite path.
The sets of all infinite and finite paths in $\dtmc$
are denoted $\ipaths_\dtmc$ and $\fpaths_\dtmc$, respectively.
We define a probability measure $\Prob_{\dtmc}$ over the set of paths $\ipaths_{\dtmc}$. 

\begin{definition}[MDP]
A \emph{Markov decision process} (MDP) is a tuple $\mdp = \mdptuple$, where
states $S$, initial state $\sinit$, atomic propositions $\AP$ and labelling $\lab$
are as for a DTMC,
$\Act$ is a finite set of \emph{actions},
and $P:S\times\Act\times S\ra[0,1]$ is a probabilistic transition function satisfying
$\forall s\in S,\forall a\in\Act: \sum_{s'\in S}P(s,a,s')\in\{0,1\}$.
\end{definition}

In each state $s$ of an MDP $\mdp$, there are one or more \emph{available} actions which can be taken,
denoted $\Act(s) = \{a\in\Act\,|\,P(s,a,s')>0\mbox{ for some } s'\}$.
If action $a$ is taken in $s$, the probability of taking a transition
from $s$ to $s'$ is $P(s,a,s')$, also denoted $P(s' |s, a)$.
Paths are defined in similar fashion to DTMCs but are now alternating sequences
of states and actions $s_0 a_0 s_1 a_1 s_2 \dots$ where 
$a_{i}\in\Act(s_i)$ and $P(s_i,a_i,s_{i+1})>0$ for all $i\geq 0$,
and the sets of all infinite and finite paths 
are $\ipaths_\mdp$ and $\fpaths_\mdp$, respectively.

The choice of actions in each state is resolved by a \emph{policy} (or \emph{strategy}),
based on the execution of the MDP so far. Formally, a policy takes the form $\policy:\fpaths\ra\Act$.
We say that $\policy$ is \emph{memoryless} if the mapping $\policy(\fpat)$
depends only on $\last(\fpat)$, the final state of $\fpat$, and \emph{finite-memory}
if it depends only on $\last(\fpat)$ and the current memory value,
selected from a finite set and updated at each step of execution.
The set of all policies for MDP $\mdp$ is denoted $\policies_\mdp$.

Under a given policy $\policy$, the resulting set of (infinite) paths has, as for DTMCs,
an associated probability measure, which we denote $\Pr_{\mdp}^{\policy}$.
Furthermore, for both memoryless and finite-memory policies,
we can build a (finite)  \emph{induced DTMC}
which is equivalent to $\mdp$ acting under $\policy$.

\begin{definition}[Reward structure]\label{rew-def}
A \emph{reward structure} is, for a DTMC $\dtmc$, a function $\rew : S \ra \Nset$ and,
for an MDP $\mdp$, a function $\rew : S\times\Act \ra \Nset$.
\end{definition}

For consistency with the literature on probabilistic model checking and temporal logics,
we use the terminology \emph{rewards} although in practice these can (and often do)
represent \emph{costs}, such as time elapsed or energy consumed.
For the purposes of our algorithms, we assume that rewards are integer-valued,
but we note that these could be defined as rationals, using appropriate scaling.
For an infinite path $\ipat$, we also write $\rew(\ipat,k)$
for the sum of the reward values over the first $k$ steps of the path, i.e.,
$\rew(s_0 s_1 s_2 \dots, k)=\sum_{i=0}^{k-1} \rew(s_i)$ for a DTMC
and $\rew(s_0 a_0 s_1 a_1 s_2 \dots, k)=\sum_{i=0}^{k-1} \rew(s_i,a_i)$ for an MDP.

To reason about rewards, we define random variables over the executions (infinite paths)
of a model, typically defined as the total reward accumulated along a path, up until some event occurs.
Formally, for a DTMC $\dtmc$, such a random variable is defined as
a function of the form $X : \ipaths_\dtmc \ra \Rset$, 
with respect to the probability measure $\Pr_\dtmc$ over $\ipaths_\dtmc$.
For an MDP $\mdp$ and policy $\policy\in\policies_\mdp$,
a random variable is defined as a function $X : \ipaths_\mdp \ra \Rset$, 
with respect to the probability measure $\Pr_\mdp^\policy$.



\section{Distributional Probabilistic Model Checking} \label{sec:theory} 
We formulate our approach as a \emph{distributional} extension of probabilistic model checking,
which is a widely used framework for formally specifying and verifying
quantitative properties of probabilistic models.
In particular, we build upon existing temporal logics in common use.
The core property we consider is the probability distribution over
the amount of reward (or cost) that has been accumulated until some specified sequence of events occurs
(which could constitute, for example, the successful completion of a task by a robot).

To represent events, we use the co-safe fragment~\cite{KV01} of linear temporal logic (LTL)~\cite{Pnu81}.
%
%
LTL formulae are evaluated over infinite paths of a model
labelled with atomic propositions from the set $\AP$ but,
for use with cumulative reward, we restrict our attention to the \emph{co-safe} fragment,
containing formulae which are satisfied in finite time.
Formally, this means any satisfying path ($\ipat\sat\psi$) has a \emph{good prefix},
i.e., a finite path prefix $\fpat'$ such that $\fpat' \ipat'' \sat \psi$ for any suffix $\ipat''$.

The key ingredient of our temporal logic specifications is a \emph{distributional query},
which gives a property (such as the expected value, or variance) of
the distribution over the accumulated reward until an event's occurrence.

\begin{definition}[Distributional query]
For a DTMC, a \emph{distributional query} takes the form
$\smash{\rewop{\rvprop(\rew)}{=?}{\psi}}$, where
$\rew$ is a reward structure,
$\rvprop$ is a random variable property (e.g., $\exp, \var, \sd, \mode, \VaR{},\CVaR{}$),
and $\psi$ is a formula in co-safe LTL.
\end{definition}
%
%
Examples of distributional queries for a DTMC are:
%

\begin{itemize}
\item
$\rewop{\var(\rew_\mathit{energy})}{=?}{\future(\mathit{goal}_1\land\future\mathit{goal}_2)}$
-- the variance in energy consumption until a robot visits location $\mathit{goal}_1$ followed by location $\mathit{goal}_2$;
\item
$\rewop{\mathrm{mode}(\rew_\mathit{coll})}{=?}{\future\mathit{sent_1}\lor\future\mathit{sent_2}}$
\-- the most likely number of packet collisions before a communication protocol successfully sends one of two messages.
\end{itemize}
For an MDP, the goal is to optimize a random variable property $\rvprop$ over its policies, which we call \emph{distributional optimization queries}.
In this paper, we focus on two particular cases,
expected value ($\E$) and conditional value-at-risk ($\CVaR{}$),

\begin{definition}[Distributional optimization query]
For an MDP, a \emph{distributional optimization query} takes the form
$\smash{\rewop{\rvprop(\rew)}{\opt=?}{\psi}}$, where
$\rew$ is a reward structure,
$\rvprop\in\{\exp,\CVaR{}\}$, 
$\opt\in\{\min,\max\}$
and $\psi$ is a formula in co-safe LTL.
%
For the resulting policy, we can perform \emph{policy evaluation}
on the induced DTMC using one or more other distributional queries $\smash{\rewop{\rvprop'(\rew')}{=?}{\psi'}}$.
\end{definition}
An example optimization query is $\rewop{\CVaR{0.9}(\rew_\mathit{time})}{\min=?}{\future\mathit{goal}}$,
which minimizes the conditional value-at-risk 
with respect to the time for a robot to reach its goal.



\startpara{Semantics}
A distributional query $\rewop{\rvprop(\rew)}{=?}{\psi}$
is evaluated on a DTMC $\dtmc$,
and a distributional optimization query $\rewop{\rvprop(\rew)}{\opt=?}{\psi}$
on an MDP $\mdp$,
in each case via a random variable for the reward accumulated from its initial state:
%
\begin{eqnarray*}
\label{eq:semdtmc}
\rewop{\rvprop(\rew)}{=?}{\psi}
& = &
\rvprop(X_\dtmc^{r,\psi}) \\
\label{eq:semmdpmin}
\rewop{\rvprop(\rew)}{\min=?}{\psi}
= 
\inf_{\policy\in\policies_\mdp} \rvprop(X_{\mdp,\policy}^{r,\psi}) & \text{ or } & \label{eq:semmdpmax}
\rewop{\rvprop(\rew)}{\max=?}{\psi}
= 
\sup_{\policy\in\policies_\mdp} \rvprop(X_{\mdp,\policy}^{r,\psi})
\end{eqnarray*}
where the random variables $X_\dtmc^{r,\psi}:\ipaths_\dtmc\ra\Rset$, $X_{\mdp,\policy}^{r,\psi}:\ipaths_\mdp\ra\Rset$
are: 
\begin{eqnarray*}\label{eq:sempath}
X_\dtmc^{r,\psi}(\ipat) \ = \ 
X_{\mdp,\policy}^{r,\psi}(\ipat) \ = \ 
\{
\rew(\omega,k_\psi-1)
\text{ if }\omega\sat\psi \text{; } 
\infty
\text{ otherwise}\}
\end{eqnarray*}
and $k_\psi = \min \{ k \mid (\ipat,k) \sat \psi \}$
is the length of the shortest good prefix for $\psi$.

\begin{examp}
We illustrate our framework with an example of
an autonomous robot navigating within a risky environment
modelled as an MDP (Figure~\ref{fig:mud_nails}).
The robot starts in the leftmost location (blue circle),
and may pass through two types of terrain, mud (orange zigzag)
and ground littered with nails (purple hatching).
%
The default cost of navigation is $1$ per step, obstacles (gray) incurring a cost of $35$.
In the ``nails'' terrain, there is a probability of $0.2$
incurring a cost of $5$;
the ``mud'' terrain is safer but slower: a fixed cost of $3$ per step.
%
\begin{figure}[!t]
\centering
\includegraphics[width=.75\linewidth]{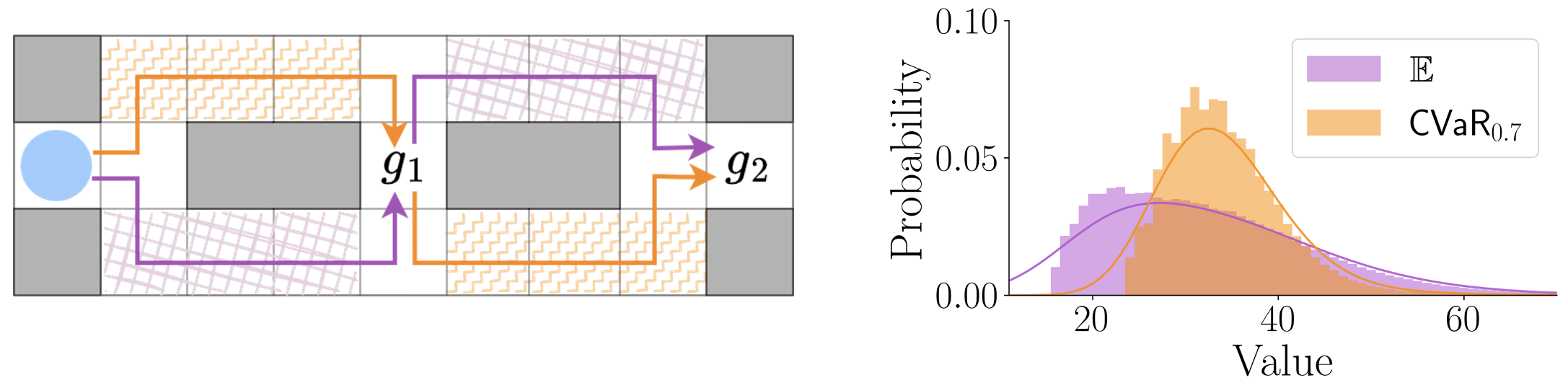}
\vspace*{-0.6em}
\caption{The ``mud \& nails'' example. Left: Map of the terrain to navigate, with two policies
that minimize expected cost and conditional value-at-risk to visit $g_1$ and then $g_2$.
Right: The corresponding distributions over cost.}
\label{fig:mud_nails}
\vspace*{-1em}
\end{figure}
%
Consider the total cost to visit $g_1$ and then $g_2$.
Given a reward structure $\mathit{cost}$ encoding the costs as above,
we can aim to minimize either the expected cost or the conditional value-at-risk, using queries
$\smash{\rewop{\E({\mathit{cost}})}{\min=?}{\future(g_1\land\future g_2)}}$
or $\smash{\rewop{\CVaR{0.7}({\mathit{cost})}}{\min=?}{\future(g_1\land\future g_2)}}$.
Figure~\ref{fig:mud_nails} also shows the resulting policies,
plotted on the map in purple and orange, respectively,
and the corresponding probability distributions over cost.
We can analyze each policy with further distributional queries,
e.g., $\smash{\rewop{\rvprop({\mathit{cost}})}{=?}{\future g_1}}$
for $\rvprop=\{\E,\var\}$
to evaluate the mean and variance of the cost to reach $g_1$.
\end{examp}

\section{Distributional Model Checking Algorithms} \label{sec:algs} 

We now describe algorithms for distributional probabilistic model checking,
i.e., to evaluate distributional queries of the form
$\rewop{\rvprop(\rew)}{=?}{\psi}$ for a DTMC
or $\rewop{\rvprop(\rew)}{\opt=?}{\psi}$ for an MDP.
Following the semantics given in Section~\ref{sec:theory},
for a DTMC $\dtmc$, this necessitates generating the probability distribution
of the random variable $X_\dtmc^{r,\psi}$,
corresponding to reward structure $\rew$ and LTL formula $\psi$, on $\dtmc$.
The value $\rvprop(X_\dtmc^{r,\psi})$
can then be evaluated on the distribution for any $\rvprop$.
For an MDP $\mdp$, we aim to find a policy $\policy^*$
which optimizes the value $\rvprop(X_{\mdp,\policy}^{r,\psi})$ over policies $\policy$.


For both classes of model, in standard fashion, we reduce the problem
to the simpler case where $\psi$ is a \emph{reachability} formula
by constructing an automaton product.
More precisely, we build a deterministic finite automaton (DFA) $\cA_\psi$
representing the ``good'' prefixes of co-safe LTL formula $\psi$,
and then construct a DTMC-DFA product $\dtmc\otimes\cA_\psi$
or MDP-DFA product $\mdp\otimes\cA_\psi$ with state space $S \times Q$,
where $S$ is the state space of the original model and $Q$ the states of the DFA.
There is a one-to-one correspondence between paths (and, for MDPs, policies) in the original model
and the product model~\cite{BK08}.
%
%
%

Hence, in what follows, we restrict our attention to computing the probability distributions
for random variables defined as the reward to reach a target set of states $T \subseteq S$,
describing first the case for a DTMC
and then the cases for risk-neutral ($\rvprop=\E$)
and risk-sensitive ($\rvprop=\CVaR{}$) optimization for an MDP.
For the latter two, for presentational simplicity, 
we focus on the case of minimization, 
but it is straightforward to adapt the algorithms to the maximizing case.


\subsection{Forward Distribution Generation for DTMCs} \label{sec:dtmc}
We fix a DTMC $\dtmc$, reward structure $\rew$ and set of target states $T$.
In this section, we describe how to compute the probability distribution
for the reward $\rew$ accumulated in $\dtmc$ until $T$ is reached,
i.e., for the random variable $\smash{X_\dtmc^{r,\futureop\,T}}$.
We denote this distribution by $\mu$.
%
Note that, since individual rewards are integer-valued,
and are summed along paths, $\mu$ is a discrete distribution. 

\setlength{\textfloatsep}{5pt}
\begin{algorithm}[!t]
\caption{Forward distribution generation for DTMCs}
\label{ag:dtmc}

\SetKwInOut{Input}{Input}
\SetKwInOut{Output}{Output}
\SetKwComment{Comment}{// }{}
\setcounter{AlgoLine}{0}

\Input{DTMC $\dtmc=\dtmctuple$, rewards $r$, target $T\subseteq S$, accuracy $\varepsilon\in\Rset_{>0}$}
\Output{The discrete probability distribution $\mu$ for $X_\dtmc^{r,\futureop\,T}$.}

$S_\infty \gets \{s\in S\ | \ s \in \mbox{BSCC $C\subseteq S$ with $C\cap T=\emptyset$}\}$;
$\mu_{\times} \gets \delta_{(\sinit,0)}$;
$p_{\overline{T}} = 1$; $p_\infty=0$ \;
\While{$p_{\overline{T}}-p_\infty > \varepsilon$}{
	$\mu_{\times}' \gets \{\}$;
	$p_{\overline{T}} \gets 0$ \;
	\For {$((s,i)\mapsto p_{s,i})\in\mu_{\times}$}{
		\If{$s \in T$}{
			$\mu_{\times}'(s,i) \gets \mu_{\times}'(s,i) + p_{s,i}$ \;
		}\Else{
			\For {$(s' \mapsto p_{s'})\in P(s,\cdot)$}{
				\If{$s'\not\in T$}{$p_{\overline{T}} \gets p_{\overline{T}} + p_{s,i} \cdot p_{s'}$ \;}
			    \If{$s'\not\in S_\infty$}{
				    $\mu_{\times}'(s',i+r(s)) \gets \mu_{\times}'(s',i+r(s)) + p_{s,i} \cdot p_{s'}$ \;
				}\Else{
				    $p_\infty \gets p_\infty + \mu_{\times}(s,i) \cdot p_{s'}$ \;
				}
			}
		}
	}
    $\mu_{\times}\gets\mu_{\times}'$ \;
}
\Return $\{i\mapsto p_i\,|\,p_i=\sum_s \mu_{\times}(s,i)\} \cup \{\infty\mapsto p_\infty\}$
\end{algorithm}

We compute the distribution in a forward manner,
up to a pre-specified accuracy $\varepsilon$, using Algorithm~\ref{ag:dtmc}.
First, note that  
the reward accumulated along
a path that never reaches the target $T$ is defined to be $\infty$ (see Section~\ref{sec:theory}).
Probabilistic model checking algorithms typically compute the \emph{expected}
reward to reach a target $T$ from a state $s$,
which is therefore infinite if $s$ has a non-zero probability of not reaching $T$.
Here, we have to take slightly more care since there may be states
from which there is a non-zero probability of both accumulating
finite and infinite reward.
This means that $\mu$ is a distribution over $\Nsetinf$.

Algorithm~\ref{ag:dtmc} first identifies the states $S_\infty$ of $\dtmc$
from which the probability of accumulating infinite reward is 1,
which are those in bottom strongly connected components (BSCCs) of $\dtmc$
that do not intersect with $T$.
It then computes a discrete distribution $\mu_{\times}$ over $S\times\Nsetinf$ where,
at the $k$th iteration, $\mu_{\times}(s,i)$ is the probability of being in state $s$
and having accumulated reward $i$ after $k$ steps.
A new version $\mu_{\times}'$ is computed at each step.
Abusing notation, we write distributions as lists $\{x_1\mapsto p_1,\dots\}$
of the elements $x_j$ of their support and their probabilities $p_j$.
We also keep track of the probabilities $p_{\overline{T}}$ and $p_\infty$
of, by the $k$th iteration, \emph{not} having reached the target set $T$
and being in $S_\infty$, respectively.
The distribution $\mu$ is finally computed by summing $\mu_{\times}(s,i)$ values over all states and can be analyzed with additional distributional properties. 


\startpara{Correctness and convergence}
Let $\mu$ be the exact distribution for $\smash{X_\dtmc^{r,\futureop\,T}}$
and $\hat{\mu}$ be the one returned by Algorithm~\ref{ag:dtmc},
using accuracy $\varepsilon>0$.
We have: 
\begin{eqnarray}\label{eq:dtmc_correctness}
\mu(i)\leq\hat{\mu}(i)\leq\mu(i){+}\varepsilon && \mbox{ for all } i\in\Nsetinf
\end{eqnarray}
Note that the support of $\mu$ may be (countably) infinite,
but $\hat{\mu}$ is finite by construction.
In this case, the total truncation error is also bounded by $\varepsilon$:
if $\hat{k}\in\Nset$ is the maximum finite value in the support of $\hat{\mu}$,
then $\sum\nolimits_{\hat{k}<i<\infty}\mu(i)\leq\varepsilon$.

To see the correctness of Equation $(\ref{eq:dtmc_correctness})$,
observe that $\hat{\mu}(i)$ is ultimately computed from the sum of
the values $\sum_{s}\mu_{\times}(s,i)$ in Algorithm~\ref{ag:dtmc},
the total value of which is non-decreasing since rewards are non-negative.
In any iteration, at most $p_{\overline{T}}-p_\infty$ will be
added to any value $\mu_{\times}(s,i)$ and, on termination, $p_{\overline{T}}-p_\infty\leq\varepsilon$.
Convergence is guaranteed for any $\varepsilon>0$:
since we separate the states $S_\infty$ in non-target BSCCs,
within $k$ iterations,
the combined probability of having reached $T$ (i.e., $1-p_{\overline{T}}$)
or reaching $S_\infty$ (i.e., $p_\infty$)
tends to $1$ as $k\ra\infty$.



\subsection{Risk-Neutral Distributional Value Iteration for MDPs} \label{sec:neutral}

In this section, we present a \emph{risk-neutral} DVI method,
for computing value distributions of states of an MDP $\mdp$ under an optimal policy
that minimizes the \emph{expected} cumulative reward to reach a target set $T\subseteq S$,
i.e., minimizes $\E(X_{\mdp,\policy}^{r,\futureop\,T})$
for random variables $\smash{X_{\mdp,\policy}^{r,\futureop\,T}}$ of MDP policies $\policy$.
%
In contrast to the case for DTMCs, 
we now assume that there exists an optimal policy with finite expected reward,
i.e., which reaches the target set $T$ with probability 1.
This can be checked efficiently with an analysis of the underlying graph of the MDP~\cite{BdA95}.

\begin{algorithm}[t]
\caption{Risk-Neutral Distributional Value Iteration}
\label{ag:neutral}
\SetKwInOut{Input}{Input}
\SetKwInOut{Output}{Output}
\SetKwFunction{proj}{proj}
\setcounter{AlgoLine}{0}
\Input{MDP $\mdp = \mdptuple$, rewards $r$, target $T\subseteq S$, and $\epsilon\in\Rset_{>0}$}
\Output{optimal policy $\pi^*$ for query $\rewop{\exp(\rew)}{\min=?}{\futureop\,T}$, distribution $\mu_{\sinit}$ under $\pi^*$}
 $e=\infty$; $\mu_s \gets \delta_0, \forall s\in S$\;
 \While{$e > \epsilon$}{
    \ForEach{ $s \in S \setminus T$}{
        \ForEach{$a \in \Act(s)$}{
            $\eta(s, a) \deq$ \proj{$ r(s,a) + \sum_{s'\in S} P(s,a,s') \cdot \mu_{s'}$} 
        }
        $\pi^*(s) \gets \argmin_{a \in \Act(s)} \exp\big( X | {X \sim \eta(s,a)} \big)$ ;
        $\mu'_s \gets \eta(s, \pi^*(s))$ \;
    }
    $e \gets \sup_{s \in S \setminus T} d(\mu_s, \mu'_s)$ \;
    $\mu_s \gets \mu'_s, \forall s\in S$\;
 }  
 \Return $\pi^*$ and $\mu_{\sinit}$
\end{algorithm}

The risk-neutral DVI method is shown in \agref{ag:neutral}.
For each MDP state $s \in S$, it initializes its value distribution
$\mu_s$ to Dirac distribution $\delta_0$. 
The algorithm loops through lines 2-8 to update value distributions of any non-target state $s \in S \setminus T$ as follows. 
For each available action $a \in A(s)$ in state $s$, a value distribution is obtained via the distributional Bellman equation shown in line 5 then projected to $\eta(s,a)$ to match the chosen representation (see Sec.~\ref{sec:dist}).
The optimal action $\pi^*(s)$ in state $s$ is the one that achieves the minimal expected value of $\eta(s,a)$.
The updated value distribution $\mu'_s$ of state $s$ is given by $\eta(s, \pi^*(s))$.
The algorithm terminates when the supremum of distributional distance $d(\mu_s, \mu'_s)$ across all states 
(the choice of metrics is discussed below)
is less than the convergence threshold $\epsilon$.
Unlike the accuracy $\varepsilon$ for Algorithm~\ref{ag:dtmc}, this threshold $\epsilon$ does \emph{not} provide a guarantee on the precision of the result after convergence (similar issues occur in classical value iteration for MDPs~\cite{HM14}).


\startpara{Distributional approximation}
To enable a practical implementation of the algorithm, we need a probability distribution representation with finitely many parameters to store value distributions in memory. 
Here, we can adopt the categorical (see \defref{def:cgr}) or quantile (see \defref{def:qtr}) representations.
Specifically, we need to apply the categorical or quantile projection (see ~\cite{bookdr2022}) after each update of the distributional Bellman equation (line 5).
We use the supremum Cram\'{e}r distance $\overline{\ell}_2$ for categorical representations and the supremum Wasserstein distance $\overline{w}_1$ for quantile representations as the distance metric in line 7 (see~\cite{bookdr2022} for distributional distance definitions).

\startpara{Policy convergence}
When there exists a unique risk-neutral optimal policy, \agref{ag:neutral} is guaranteed to converge to it (following~\cite[Theorem 7.9]{bookdr2022}).
However, when there are multiple optimal policies, risk-neutral DVI may fail to converge (see~\cite[Section 7.5]{bookdr2022}).
Furthermore, inaccuracies due the use of distributional approximations
could potentially lead to a sub-optimal policy being chosen.
To mitigate this, for either categorical or quantile representations,
increasing the number $m$ of atoms used yields tighter approximation error bounds~\cite{bookdr2022}.



\vspace*{-0.3cm}
\subsection{Risk-Sensitive Distributional Value Iteration for MDPs} \label{sec:cvar}

By contrast to risk-neutral policies that seek to minimize the expected reward, \emph{risk-sensitive} policies make decisions accounting for risk properties. 
In this section, we present a risk-sensitive DVI method
for minimizing the \emph{conditional value-at-risk} of reaching a target set in an MDP $\mdp$,
i.e., minimizing $\CVaR{\alpha}(X_{\mdp,\policy}^{r,\futureop\,T})$
for random variables $X_{\mdp,\policy}^{r,\futureop\,T}$ of MDP policies $\policy$.
%
Our method follows a key insight from~\cite{bauerle2011markov,rockafellar2002conditional} that conditional value-at-risk can be represented as the solution of a convex optimization problem. 

\begin{lemma}[Dual Representation of $\CVaR{}$~\cite{bauerle2011markov,rockafellar2002conditional}] 
\label{lem:cvar-duel}
Let $[x]^+$ denote the function that is 0 if $x<0$, and $x$ otherwise. 
Given a random variable $X$ over the probability space $(\Omega, \cF, \Pr)$, it holds that:
\begin{eqnarray}\label{eqn:cvar_duel}
\CVaR{\alpha}(X) = \min_{b\in\mathbb{R}} \left\{ b + \frac{1}{1-\alpha} \exp\big(\left[X - b \right]^+ \big) \right\}, 
\end{eqnarray}
and the minimum-point is given by $b^* = \VaR{\alpha}(X)$.
\qed
\end{lemma}

Intuitively, the \emph{slack variable} $b \in [\Vmin, \Vmax]$ encodes the risk budget and possible $\VaR{\alpha}(X)$ values. Since $\VaR{\alpha}(X) \in [\Vmin, \Vmax] $, the slack variable is similarly bounded by the minimum and maximum possible accumulated reward within the MDP, respectively.
We assume that the reward values are bounded and the probability of reaching the target states is $1$, therefore $\Vmin$ and $\Vmax$ are also bounded. 
To enable efficient computation, we consider a discrete number of values for $b$.
More precisely, we define a set $B$ with $n$ evenly-spaced atoms $b_1 < \cdots < b_n$ such that 
$b_1=\Vmin$, $b_n=\Vmax$, and the stride between two successive atoms is $\varsigma_n=\frac{\Vmax-\Vmin}{n-1}$.
Based on \lemref{lem:cvar-duel}, determining the optimal slack variable value $b^*$ requires computation of $\VaR{\alpha}$ for the distribution, which cannot be obtained \emph{a priori}. 
Thus, we consider all possible risk budgets.

\begin{algorithm}[t]
\caption{Risk-Sensitive Distributional Value Iteration}
\label{ag:cvar}
\SetKwInOut{Input}{Input}
\SetKwInOut{Output}{Output}
\SetKwFunction{proj}{proj}
\setcounter{AlgoLine}{0}
\Input{MDP $\mdp = \mdptuple$, reward structure $r$, target set $T\subseteq S$, risk level $\alpha$, slack variable set $B$, convergence threshold  $\epsilon\in\Rset_{>0}$}
\Output{optimal policy $\pi^*$ for query $\rewop{\CVaR{\alpha}(\rew)}{\min=?}{\futureop\,T}$, distribution $\mu_{\sinit}$ under $\pi^*$}
Construct product MDP $\mdpb=\mdpbtuple$ \;  
$\mu_{\saug{s}{b}} \gets \delta_0$,  $\forall \saug{s}{b} \in S \times B$\;
\While{$e > \epsilon$}{
    \ForEach{$\saug{s}{b} \in (S \setminus T) \times B$}{
        \ForEach{$a \in \Act(s)$}{
            $\eta(\saug{s}{b}, a) \deq $ \proj{$r(s,a) + \sum_{\saug{s'}{b'}\in S \times B} P^\mathsf{b}(\saug{s}{b}, a, \saug{s'}{b'}) \cdot \mu_{\saug{s'}{b'}}$}
        }
        $\pi^{\mathsf{b}}(\saug{s}{b}) \gets \argmin_{a \in \Act(s)} \exp \big(\left[X - b \right]^+ | {X \sim \eta(\saug{s}{b}, a)} \big)$ \;
        $\mu'_{\saug{s}{b}} \gets \eta(\saug{s}{b}, \pi^{\mathsf{b}}(\saug{s}{b}))$ \;
    }
    $e \gets \sup_{\saug{s}{b} \in (S \setminus T) \times B} d(\mu_{\saug{s}{b}}, \mu'_{\saug{s}{b}})$ \;
    $\mu_{\saug{s}{b}} \gets \mu'_{\saug{s}{b}}$, $\forall \saug{s}{b} \in (S \setminus T) \times B$\;  
}
$\binit^* \gets \argmin_{\binit \in B} \CVaR{\alpha}(X | X \sim \mu_{\saug{\sinit}{\binit}}), \forall \binit \in B$ \;
$\pi^* \gets$ policy $\pi^{\mathsf{b}}$ of the product MDP $\mdpb$ with initial state fixed to ${\saug{\sinit}{\binit^*}}$\;
\Return $\pi^*$ and $\mu_{\saug{\sinit}{\binit^*}}$
\end{algorithm}

\agref{ag:cvar} illustrates the proposed method.
We construct a product MDP model $\mdpb=\mdpbtuple$.
Unlike the product MDP defined in \sectref{sec:theory}, this MDP has multiple initial states,
one state $\saug{\sinit}{\binit}$ for each risk budget $\binit \in B$,
where $\sinit$ is the initial state of the MDP $\mdp$.
For each transition $s \xrightarrow{a} s'$ in $\mdp$ with $P(s,a,s')>0$, 
there is a corresponding transition $\saug{s}{b} \xrightarrow{a} \saug{s'}{b'}$ in $\mdpb$, where $b'$ 
is obtained by rounding down the value of $b-r(s,a)$ to the nearest smaller atom in $B$ 
and $P^\mathsf{b}(\saug{s}{b}, a, \saug{s'}{b'}) = P(s,a,s')$.
The labelling function is given by $L^\mathsf{b}(\saug{s}{b})=L(s)$. 
Next, in lines 2-12, \agref{ag:cvar} initializes and updates the value distribution of 
each augmented state $\saug{s}{b} \in S \times B$ in the product MDP $\mdpb$ in a similar fashion to the risk-neutral DVI described in \sectref{sec:neutral}. 
However, when choosing the optimal action (line 8), \agref{ag:cvar} adopts a different criterion that minimizes $\exp(\left[X - b \right]^+)$ based on the dual representation of $\CVaR{}$ (see \eqnref{eqn:cvar_duel}).

Different choices of the initial risk budget $\binit$ lead to various value distributions.
Once DVI on the product MDP $\mdpb$ converges, the algorithm selects the optimal risk budget, denoted by $\binit^*$, that yields the minimum $\CVaR{}$ of all possible initial value distributions $\mu_{\saug{\sinit}{\binit}}$. 
Finally, the algorithm returns the optimal policy $\pi^*$ 
resulting from the risk-sensitive DVI on the product MDP $\mdpb$ with initial state ${\saug{\sinit}{\binit^*}}$,
and returns the distribution $\mu_{\saug{\sinit}{\binit^*}}$.


\startpara{Correctness and convergence}
Following~\cite[Theorem 3.6]{bauerle2011markov}, when the slack variable $b$ is continuous (i.e., $B=\Rset$), there exists a solution $b^*$ of \eqnref{eqn:cvar_duel} and the optimal policy $\pi^{\mathsf{b}}$ of product MDP $\mdpb$ with initial state fixed to $\saug{\sinit}{b^*}$ is the $\CVaR{}$ optimal policy of MDP $\mdp$. 
\agref{ag:cvar}, which uses a discretized slack variable (i.e., the set of atoms $B$ is finite), converges to the same optimal policy $\pi^{\mathsf{b}}$ as $|B|$ increases, which is formalised below
\ifthenelse{\isundefined{\techreport}}{%
(and a proof can be found in the extended version of this paper~\cite{dpmc-arxiv}).
}{%
(and a proof can be found in the appendix).
}

\begin{lemma}
\label{lem:slack_variable}
    Let $\pi_1$ denote the optimal policy for minimizing $\CVaR{\alpha}(X_{\mdp,\policy}^{r,\futureop\,T})$, which is obtained with a continuous slack variable.
    Let $\pi_2$ denote the optimal policy returned by \agref{ag:cvar} where $B$ is a finite set of $n$ evenly-spaced atoms with stride $\varsigma_n$. It holds that
    $\CVaR{\alpha}(X_{\mdp,\pi_2}^{r,\futureop\,T}) - \CVaR{\alpha}(X_{\mdp,\pi_1}^{r,\futureop\,T}) = \cO(\varsigma_n)$. As $\varsigma_n$ tends to 0 (i.e., $|B|$ increases), $\pi_2$ converges to the $\CVaR{}$ optimal policy.
\qed
\end{lemma}

\section{Experiments} \label{sec:exp}
We built and evaluated a prototype implementation\footnote{Code and models are at \href{https://www.prismmodelchecker.org/files/nfm24dpmc}{https://www.prismmodelchecker.org/files/nfm24dpmc}.}
of our distributional probabilistic model checking approach 
based on PRISM~\cite{KNP11}, extending its Java explicit-state engine.
%
Our evaluation focuses initially on solving MDPs using the DVI methods (of Sections~\ref{sec:neutral} and \ref{sec:cvar}), then on solving the resulting policies using the DTMC method (of \sectref{sec:dtmc}).
All experiments were run on a machine with an AMD Ryzen 7 CPU and 14 GB of RAM allocated to the JVM.
We set $\Vmin=0$ for all case studies; $\Vmax$ varies, as detailed below.


\vspace*{-0.3em}
\subsection{Case Studies} \label{sec:studies}
\vspace*{-0.2em}

\startpara{Betting Game}
This case study is taken from~\cite{rigter2022planning}. 
The MDP models an agent with an amount of money, initially set to $5$,
which can repeatedly place a bet of amount $0\leq\lambda\leq 5$. 
The probability of winning is 0.7, the probability of losing is 0.25, and the probability of hitting a jackpot (winning $10\lambda$) is 0.05.
The game ends after 10 stages.
The reward function is given by the maximal allowance (e.g., 100) minus the final amount of money that the agent owns.
We use $\Vmax=100$.

\startpara{Deep Sea Treasure}
This case study is also taken from~\cite{rigter2022planning}.
The model represents a submarine exploring an area to collect one of several treasures. 
At each time step, the agent chooses to move to a neighbouring location;
it succeeds with probability 0.6, otherwise moves to another adjacent location with probability 0.2.
The agent stops when it finds a treasure or has explored for 15 steps. 
The reward function is defined based on the travel cost (5 per step) and opportunity cost (i.e., maximal treasure minus collected treasure value). 
We set $\Vmax=800$.

\startpara{Obstacle}
This case study is inspired by the navigation example in~\cite{chow2015risk}.
We consider an MDP model of an $N \times N$ gridworld with a set of scattered obstacles. 
The agent's goal is to navigate to a destination, while avoiding obstacles which cause a delay.  
At each time step, the agent moves in a selected direction with probability 0.9 and an unintended direction with probability 0.1.
The reward function is given by the time spent to reach the destination. 
We use $\Vmax=600$.

\startpara{UAV}
This case study is adapted from the MDP model of the interaction between
a human and an unmanned aerial vehicle (UAV) from~\cite{feng2016synthesis}.
A UAV performs road network surveillance missions with the assistance of a human operator,
and is given a mission specified with LTL formula 
$\psi = (\future w_2) \land (\future  w_5) \land (\future w_6$), which translates into covering waypoints $w_2$, $w_5$ and $w_6$ in any order.
The reward function is given by the mission completion time. 
We pick $\Vmax=500$.

\startpara{Energy}
This case study considers a robot navigating an $N \times N$ gridworld with energy constraints.
At each time step, the robot moves to an adjacent grid location with probability 0.7, or to an unintended adjacent location otherwise. 
It starts with a fixed amount of energy and consumes 1 unit per step.
The robot can only recharge its battery in the charging station. 
When the energy is depleted, the robot is transported with a delay to the charging station.
The robot is asked to complete a mission specified with LTL formula $\psi = (\future w_1) \land (\future  w_2) \land (\future w_3$).
The reward function represents the mission completion time. 
We use $\Vmax=500$.



\setlength{\tabcolsep}{0.225em} 

\begin{table}[t]
\caption{Experimental results: Timing and accuracy of each method.}
\label{tab:results_error}
\begin{tabular}{@{}llc|ccc|ccc@{}}
\toprule
\multicolumn{1}{c}{Model} & \multicolumn{1}{c}{Method} & MDP & Time (s) & $\exp$ & $\CVaR{\alpha}$ & Time$_\mathsf{dtmc}$ (s) & $\Delta^\%_\exp$ & $\Delta^\%_\CVaR{}$ \\ \midrule
\multirow{3}{*}{\begin{tabular}[c]{@{}l@{}}Betting \\ Game\end{tabular}} & risk-neut. VI & $8.9 \cdot 10^2$ & $< 1$ & \textbf{61.9} & - & $< 1$ & - & - \\
 & risk-neut. DVI & $8.9 \cdot 10^2$ & $< 1$ & \textbf{61.9} & 98.0 & $< 1$ & 0.0 & 0.0 \\
 & risk-sens. DVI & $9.0 \cdot 10^4$ & 36 & 85.3 & \textbf{92.2} & $< 1$ & 0.0 & 0.0 \\ \midrule
\multirow{3}{*}{\begin{tabular}[c]{@{}l@{}}DS \\ Treasure\end{tabular}} & risk-neut. VI & $1.2 \cdot 10^3$ & $< 1$ & \textbf{359.3} & - & $< 1$ & - & - \\
 & risk-neut. DVI & $1.2 \cdot 10^3$ & $< 1$ & \textbf{359.3} & 474.6 & $< 1$ & 0.0 & 0.33 \\
 & risk-sens. DVI & $1.2 \cdot 10^5$ & 72 & 370.1 & \textbf{458.6} & $< 1$ & 0.0 & 0.32 \\ \midrule
\multirow{3}{*}{\begin{tabular}[c]{@{}l@{}}Obstacle\\ ($N=150$)\end{tabular}} & risk-neut. VI & $2.3 \cdot 10^4$ & $< 1$ & \textbf{402.8} & - & 1,838 & - & - \\
 & risk-neut. DVI & $2.3 \cdot 10^4$ & 97 & \textbf{402.7} & 479.2 & 1,838 & 0.01 & 1.95 \\
 & risk-sens. DVI & $2.3 \cdot 10^6$ & 15,051 & 402.9 & \textbf{478.4} & 1,673 & 0.01 & 2.00 \\ \midrule
\multirow{3}{*}{UAV} & risk-neut. VI & $1.7 \cdot 10^4$ & $< 1$ & \textbf{124.1} & - & $< 1$ & - & - \\
 & risk-neut. DVI & $1.7 \cdot 10^4$ & 4 & \textbf{123.8} & 168.8 & $< 1$ & 0.2 & 0.47 \\
 & risk-sens. DVI & $1.7 \cdot 10^6$ & 2,366 & 134.9 & \textbf{169.1} & $< 1$ & 0.0 & 0.01 \\ \midrule
\multirow{3}{*}{\begin{tabular}[c]{@{}l@{}}Energy \\ ($N=15$)\end{tabular}} & risk-neut. VI & $2.6 \cdot 10^4$ & 10 & \textbf{184.3} & - & 251 & - & - \\
 & risk-neut. DVI & $2.6 \cdot 10^4$ & 108 & \textbf{184.0} & 382.0 & 234 & 0.17 & 0.47 \\
 & risk-sens. DVI & $1.3 \cdot 10^6$ & 9,384 & 184.6 & \textbf{380.9} & 122 & 0.16 & 0.33 \\ \bottomrule
\end{tabular}
\end{table}

\vspace*{-1em}
\subsection{Results Analysis} \label{sec:results}

\vspace*{-0.7em}
\startpara{Method comparison}
\tabref{tab:results_error} summarizes our experimental results across the benchmarks described above.
For each MDP, we run both the risk-neutral and risk-sensitive
variants of distributional value iteration (DVI),
optimizing expected value and $\CVaR{}$,
as described in \sectref{sec:neutral} and \sectref{sec:cvar}, respectively.
For the risk neutral case we also run standard value iteration (VI),
as implemented in PRISM.
For all three methods, we then evaluate the resulting policy,
computing the full reward distribution
using the forward distribution generation method described in \sectref{sec:dtmc}, allowing us to compute more precise results for
the expected value and $\CVaR{}$ on those policies.

The table shows the time to run each algorithm and 
the values computed during optimization
(the value for the objective being optimized is shown in bold).
Additionally, the table shows the time to run the forward distribution method on the induced DTMC, and the (percentage) relative error when comparing the VI/DVI results with the forward distribution outcomes.

For each case study, we also report the number of states in the (product) MDP that is solved.
The UAV and Energy benchmarks use non-trivial co-safe LTL formulae for the mission specification
(the others are reachability specifications) and so the MDP is a MDP-DFA product.
For risk-sensitive DVI, the state space is also augmented with a slack variable resulting in larger product MDPs.
We set the slack variable size to $|B|=51$ for the Energy model, and $|B|=101$ for the rest. 
We use the the categorical representation with $m=201$ for DVI, with $\epsilon = 0.01$ for the convergence metric. For policy evaluation, we use precision $\varepsilon = 10^{-3}$ for the Obstacle and Energy case studies and $\varepsilon= 10^{-5}$ for the others. 


\vskip6pt
\emph{Our DVI methods successfully optimize their respective objectives on a range of large MDPs.}
Generally, the policy resulting from the risk-neutral method has a lower expected value,
while the policy from the risk-sensitive method has a lower $\CVaR{\alpha}$,
and the risk-neutral method yields the same optimal policy as baseline VI.
As expected, DVI methods are more expensive than VI,
since they work with distributions, 
but the DVI methods are successfully applied to MDPs with several million states.
Additionally, the baseline VI method can only provide expected reward values,
while the distribution returned by our methods can be used to compute additional distributional properties (variance, $\VaR{}$, etc.).
Comparing the two variants of DVI, the risk-sensitive version takes considerably longer to run. This is primarily due to the use of a larger product model, incorporating a slack variable, rather than the computation required for DVI itself.
For the same reason, risk-neutral DVI scales to larger models,
but for clarity \tabref{tab:results_error} only includes models that all methods can solve.

The DTMC forward computation also works on all models.
It is often very fast (under a second in 3 cases),
but grows expensive on models where the support of the distribution is large.
From its results, we see that both DVI methods produce
approximate distributions that are close to the true distribution.


Note that in the last three case studies, the $\Vmax$ value is higher, resulting in a larger stride and thus more coarse representations for both the value distributions and the slack variable (for risk-sensitive DVI). 
This results in more approximation errors when computing metrics from the value distributions generated using DVI. 
This can be seen in the case of the UAV model where the risk-neutral method underestimates $\CVaR{\alpha}$ (168.8 compared to 169.6 from the true distribution generated by the DTMC method for the same policy).
The following experiments aim to evaluate how the parameters of the distributional representation affect the resulting approximate distributions generated by DVI.

\setlength{\textfloatsep}{20pt}
\setlength{\floatsep}{16pt}

\begin{figure}[t]
\centering
\subfloat[Categorical representation]{
  \includegraphics[width=.33\linewidth]{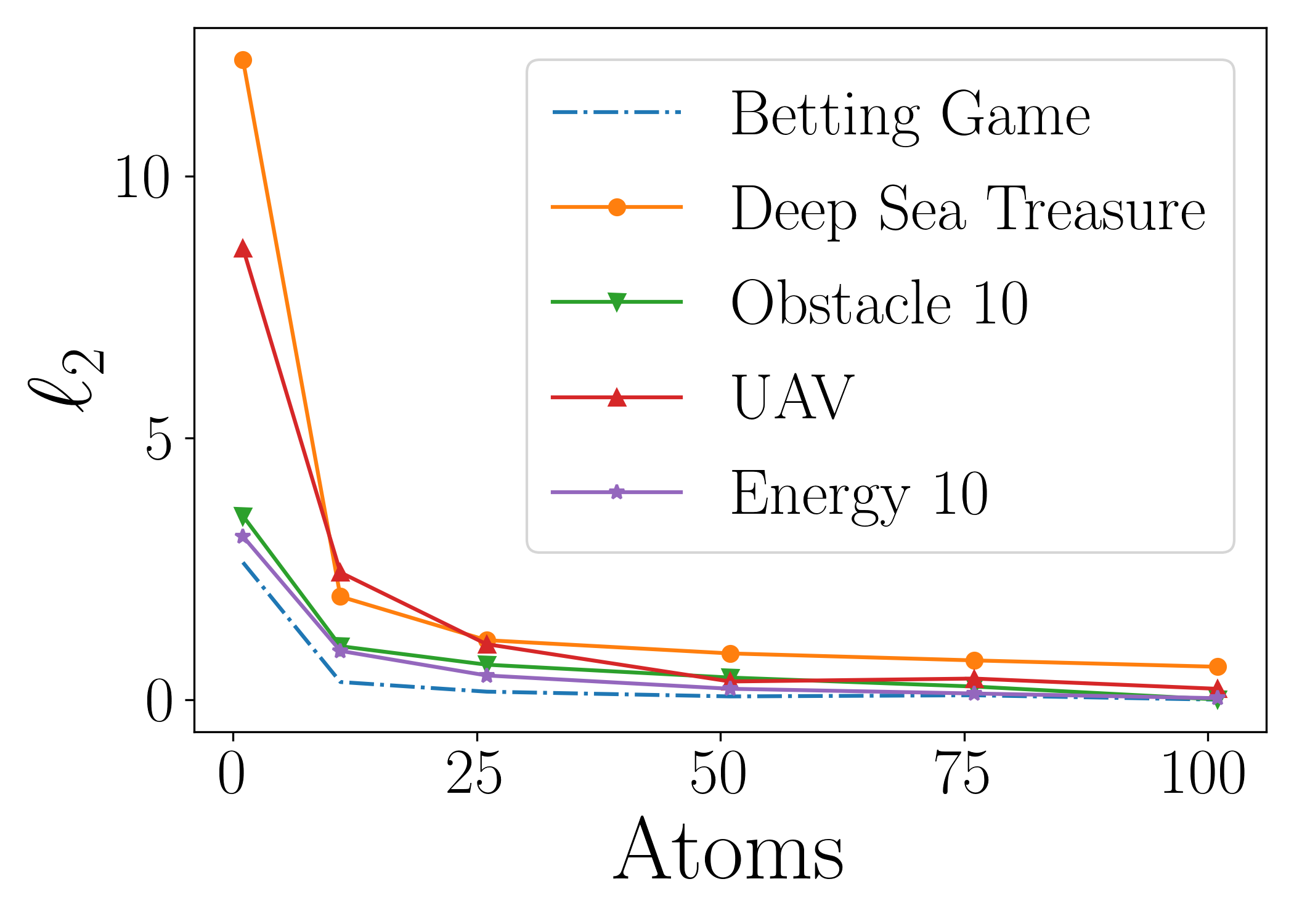}
  \label{fig:l2_graph_c51}
}
\subfloat [Quantile representation]{
  \includegraphics[width=.33\linewidth]{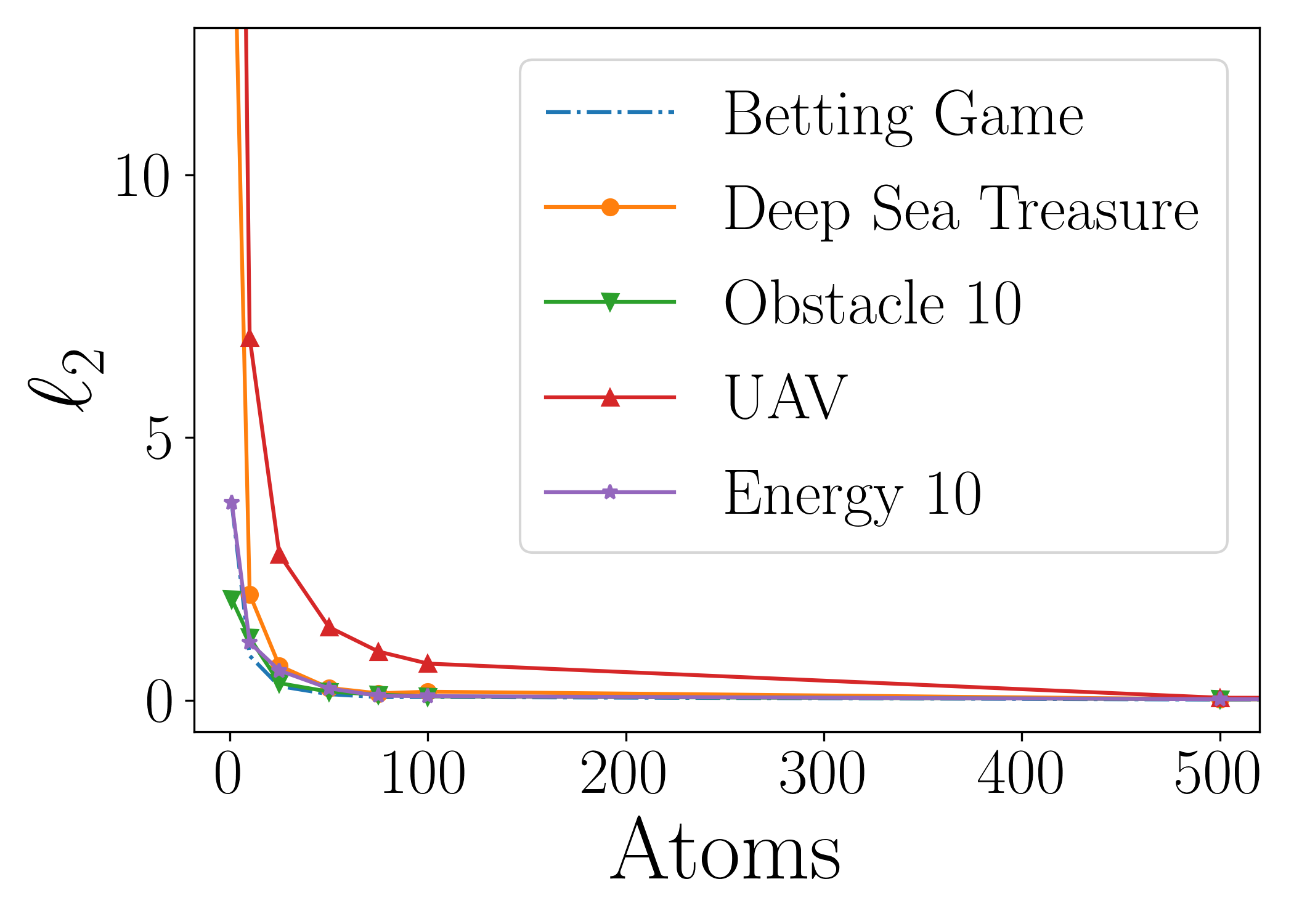}
  \label{fig:l2_graph_qr}
}
\caption{Varying the numbers of atoms for distributional representations.}
\label{fig:l2_graphs}
\vspace*{-0.8em}
\end{figure}

\startpara{Effects on distributional approximation}
\figref{fig:l2_graphs} 
plots the effects of varying the number of atoms used for categorical and quantile representations for distributions, in terms of the $\ell_2$ distance between the approximate distribution resulting from risk-neutral DVI and the ground truth (obtained via applying the DTMC forward distribution generation method with $\varepsilon=10^{-5}$ on the resulting optimal policy). 
\emph{For both representations, the $\ell_2$ distance approaches 0 as the number of atoms increases, indicating that the approximate distributions become very close to the ground truth.}
We observe similar effects with the risk-sensitive method and thus omit the resulting plot.
Note that the Deep Sea Treasure model has a larger $\Vmax$ and thus the resulting $\ell_2$ is higher than other models when using a maximum of 101 atoms in the categorical representation.    

A larger number of atoms ($m$ value) leads to a higher computational cost,
thus we consider smaller models for the Obstacle and Energy case studies with $N=10$ for plotting.
As an illustration of accuracy/cost trade-off, 
for Energy 10, 
the runtime using categorical representations with 11 atoms (resp. 101 atoms) is 0.3s (resp. 0.63s),
while the runtime when using quantile representations with 10 atoms (resp. 100 atoms) is 0.9s (resp. 5s). 
The quantile projection is more expensive than the categorical projection, resulting in higher runtimes. 

\begin{figure}[t]
\centering
\hspace{-1 em}
\subfloat [Deep Sea Treasure]{
  \includegraphics[width=.33\linewidth]{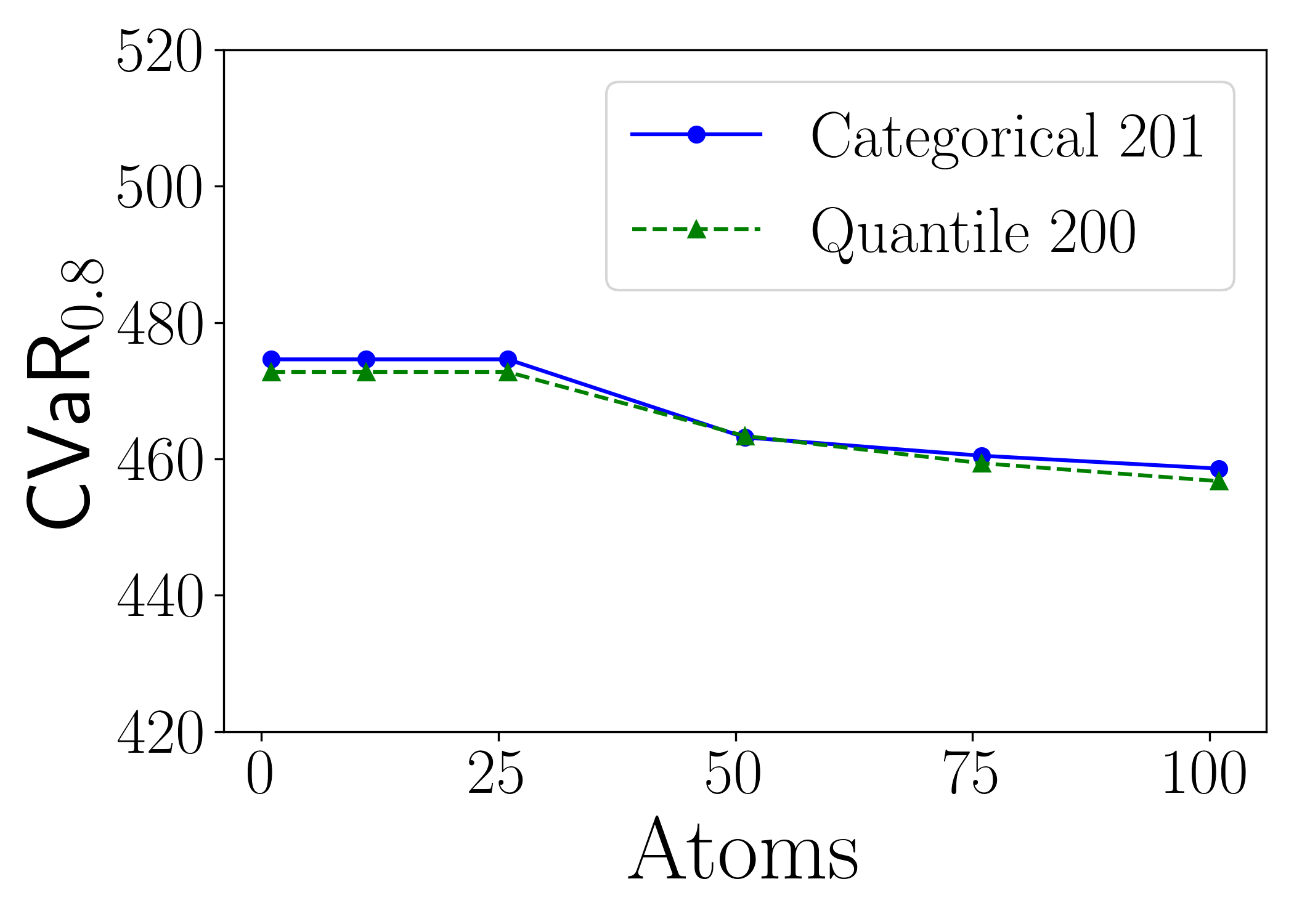}
  \label{fig:batoms_graphs_dst}
} \hspace{-1em}%
\subfloat [Obstacle 10]{
  \includegraphics[width=.33\linewidth]{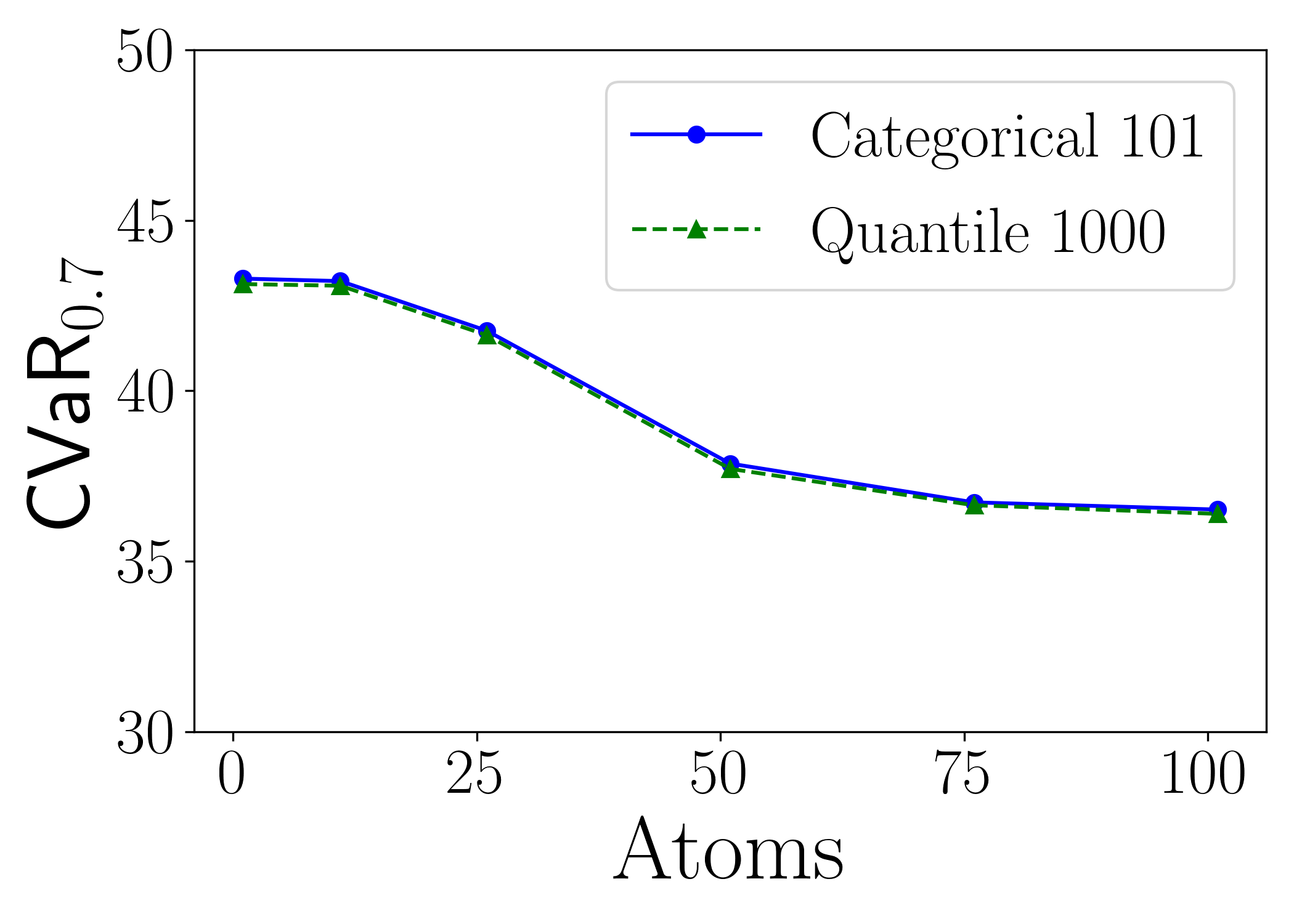}
  \label{fig:batoms_graphs_obstacle}
} \hspace{-1 em}%
\subfloat [Energy 10]{
  \includegraphics[width=.33\linewidth]{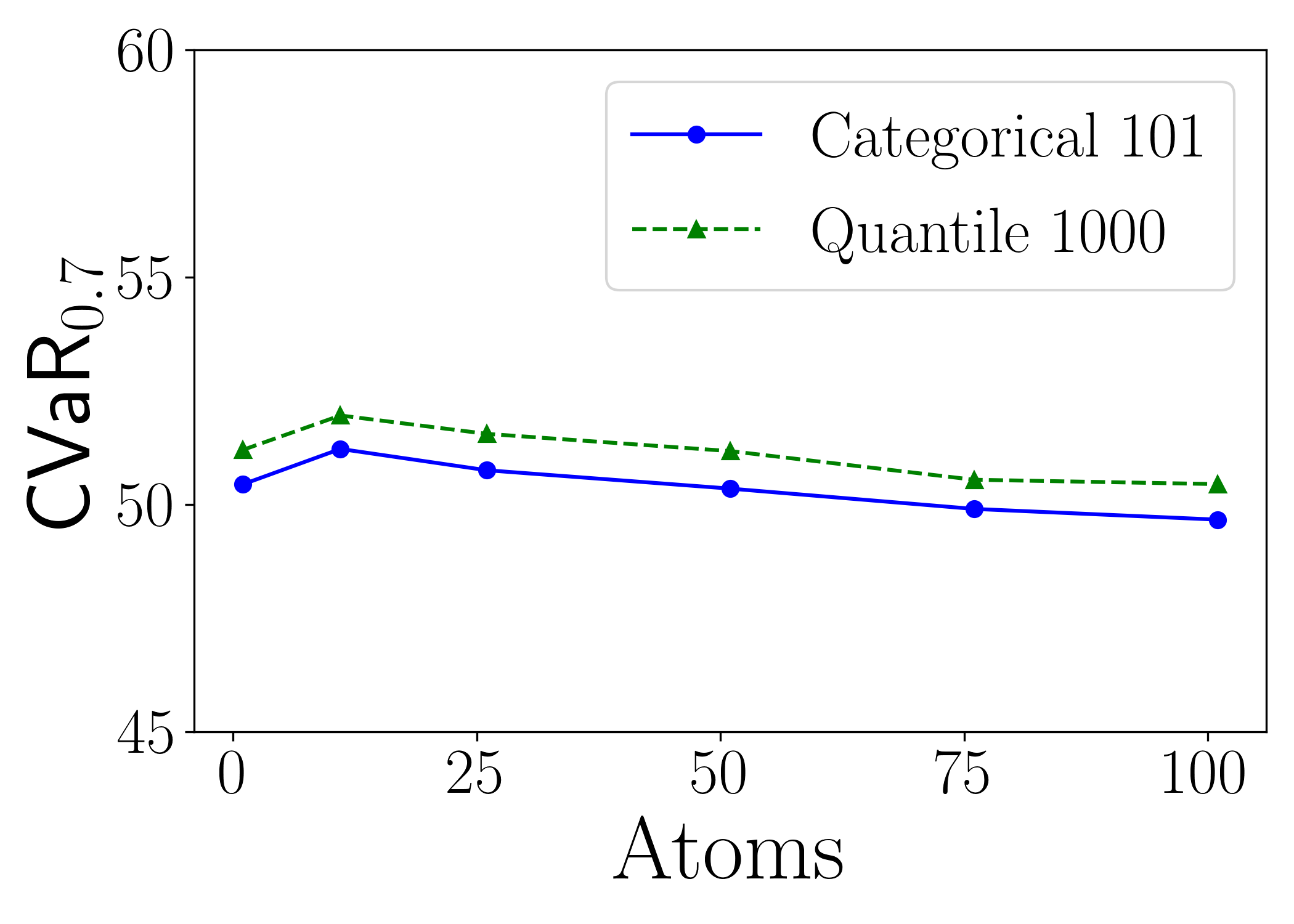}
  \label{fig:batoms_graphs_energy}
}%
\caption{Results for varying the numbers of atoms in risk-sensitive DVI.}
\label{fig:batoms_graphs}
\vspace*{-0.8em}
\end{figure}

\startpara{Effects of slack variable atoms}
\figref{fig:batoms_graphs} illustrates the effects of varying the number of atoms used for the slack variables ($|B|$) in risk-sensitive DVI. 
\emph{The results show that increasing $|B|$ generally leads to better policies with smaller $\CVaR{}$ values. }
This is in part because the algorithm would check a larger set of initial risk budgets $\binit \in B$.
But there is a trade-off since the computational cost grows with an increasing $|B|$. 
For example, in the Energy 10 model, the runtime using the categorical representation with 101 atoms for 
$|B|=11$ (resp. $|B|=101$) is 7.8s (resp. 78.6s),
whereas the runtime of using the quantile representation with 1,000 atoms for
$|B|=11$ (resp. $|B|=101$) is 477s (resp. 5,163s).


\startpara{DTMC forward computation}
Finally, we further evaluate the forward computation method for DTMCs from \sectref{sec:dtmc}
on a range of common DTMC benchmarks from the PRISM benchmark suite~\cite{KNP12b}.
In particular, we compare to an alternative computation using the risk-neutral DVI method method of \sectref{sec:neutral}, treating DTMCs as a special case of MDPs.
\tabref{tab:dtmc_compare} shows the performance of the two methods.
For each model, we indicate the parameters used for the benchmark and the DTMC size (states and transitions).
For the DVI method, we use the categorical representation with a stride of 1 and a value of $\Vmax$ large enough to represent the distribution (also shown in the table).

\textit{In two of the three models, the DTMC computation is much faster.}
This is because the DVI method calculates a reward distribution for every state.
For the third model, where $\Vmax$ is significantly higher, DVI is actually faster
(the same can be seen for the Obstacle and Energy models in \tabref{tab:results_error}).
The DTMC method computes distribution to a pre-specified accuracy,
but DVI may incur approximation errors, primarily due to convergence.
\tabref{tab:dtmc_compare} also shows (relative) errors for the expected value and $\CVaR{}$ metrics for each benchmark.

\setlength{\tabcolsep}{0.46em} 

\begin{table}[!t]
\caption{Performance comparison for DTMC forward computation}
\begin{tabular}{@{}ll|lc|cc|c|cc@{}}
\toprule
Model & Param.s & States & Transitions & $\Vmax$ & DVI(s) & DTMC(s) & $\Delta^\%_\exp$ & $\Delta^\%_{\CVaR{}}$ \\ \midrule
EGL & \multicolumn{1}{l|}{$\scriptstyle\mathsf{N=8, L=3}$} & $5.4 \cdot 10^6$ & \multicolumn{1}{c|}{$5.5 \cdot 10^6$} & 40 & 439 & \multicolumn{1}{c|}{1} & 0.4 & 0.5 \\
 & \multicolumn{1}{l|}{$\scriptstyle\mathsf{N=8, L=4}$} & $7.5 \cdot 10^6$ & \multicolumn{1}{c|}{$7.6 \cdot 10^6$} & 40 & 897 & \multicolumn{1}{c|}{1} & 0.4 & 0.4 \\
 & $\scriptstyle\mathsf{N=8, L=5}$ & $9.6 \cdot 10^6$ & $9.7 \cdot 10^6$ & 50 & 4,345 & 1 & 0.3 & 0.4 \\ 
 \midrule
Leader & \multicolumn{1}{l|}{$\scriptstyle\mathsf{N=8, K=5}$} & $2.7 \cdot 10^6$ & \multicolumn{1}{c|}{$3.1 \cdot 10^6$} & 20 & 41 & \multicolumn{1}{c|}{2} & 0.0 & 0.0 \\
 & \multicolumn{1}{l|}{$\scriptstyle\mathsf{N=10, K=4}$} & $9.4 \cdot 10^6$ & \multicolumn{1}{c|}{$1.0 \cdot 10^7$} & 30 & 577 & \multicolumn{1}{c|}{15} & 0.3 & 0.6 \\
 & \multicolumn{1}{l|}{$\scriptstyle\mathsf{N=8, K=6}$} & $1.2 \cdot 10^7$ & \multicolumn{1}{c|}{$1.3 \cdot 10^7$} & 20 & 163 & \multicolumn{1}{c|}{9} & 0.0 & 0.1 \\
\midrule
 Herman & \multicolumn{1}{l|}{$\scriptstyle\mathsf{N=13}$} & $8.2 \cdot 10^3$ & \multicolumn{1}{c|}{$1.6 \cdot 10^6$} & 100 & 4 & \multicolumn{1}{c|}{14} & 0.6 & 0.9 \\
 & \multicolumn{1}{l|}{$\scriptstyle\mathsf{N=15}$} & $3.3\cdot 10^4$ & \multicolumn{1}{c|}{$1.4 \cdot 10^7$} & 120 & 57 & \multicolumn{1}{c|}{190} & 0.5 & 0.9 \\
 & \multicolumn{1}{l|}{$\scriptstyle\mathsf{N=17}$} & $1.3 \cdot 10^5$ & \multicolumn{1}{c|}{$1.3 \cdot 10^8$} & 140 & 1,234 & \multicolumn{1}{c|}{2,369} & 0.8 & 1.2 \\
 \bottomrule
\end{tabular}
\label{tab:dtmc_compare}
\vspace*{-0.8em}
\end{table}



\section{Conclusion} \label{sec:conclu}
This paper presents a distributional approach to probabilistic model checking, which supports a rich set of distributional queries for DTMCs and MDPs. 
Experiments on a range of benchmark case studies demonstrate that our approach can be successfully applied to check various distributional properties (e.g., $\CVaR{}$, $\VaR{}$, variances) of large MDP and DTMC models.
We believe that this work paves the way for applying distributional probabilistic model checking in many safety-critical and risk-averse domains. 
For future work, we will explore distributional queries with multiple objectives and under multi-agent environments.

\startpara{Acknowledgements}
This work was supported in part by NSF grant CCF-1942836 and the ERC under the European Union’s Horizon 2020 research and innovation programme (FUN2MODEL, grant agreement No. 834115).

%

\bibliographystyle{splncs04}

\ifthenelse{\isundefined{\techreport}}{%
\bibliography{references}

\begin{thebibliography}{}
\providecommand{\url}[1]{\texttt{#1}}
\providecommand{\urlprefix}{URL }
\providecommand{\doi}[1]{https://doi.org/#1}

\end{thebibliography}


\begin{thebibliography}{10}
  \providecommand{\url}[1]{\texttt{#1}}
  \providecommand{\urlprefix}{URL }
  \providecommand{\doi}[1]{https://doi.org/#1}
  
  \bibitem{BK08}
  Baier, C., Katoen, J.P.: Principles of Model Checking. MIT Press (2008)
  
  \bibitem{bauerle2011markov}
  B{\"a}uerle, N., Ott, J.: Markov decision processes with average-value-at-risk criteria. Mathematical Methods of Operations Research  \textbf{74}(3),  361--379 (2011)
  
  \bibitem{bookdr2022}
  Bellemare, M.G., Dabney, W., Rowland, M.: Distributional Reinforcement Learning. MIT Press (2023), \url{http://www.distributional-rl.org}
  
  \bibitem{BdA95}
  Bianco, A., de~Alfaro, L.: Model checking of probabilistic and nondeterministic systems. In: Proc. 15th Conference on Foundations of Software Technology and Theoretical Computer Science (FSTTCS'95). LNCS, vol.~1026, pp. 499--513. Springer (1995)
  
  \bibitem{borkar2014risk}
  Borkar, V., Jain, R.: Risk-constrained {Markov} decision processes. IEEE Transactions on Automatic Control  \textbf{59}(9),  2574--2579 (2014)
  
  \bibitem{BCFK17}
  Brazdil, T., Chatterjee, K., Forejt, V., Kucera, A.: Trading performance for stability in {Markov} decision processes. Journal of Computer and System Sciences  \textbf{84},  144--170 (2017)
  
  \bibitem{CJJW22}
  Chen, M., Katoen, J., Klinkenberg, L., Winkler, T.: Does a program yield the right distribution? - verifying probabilistic programs via generating functions. In: Proc. 34th International Conference on Computer Aided Verification (CAV'22). LNCS, vol. 13371, pp. 79--101. Springer (2022)
  
  \bibitem{chow2014algorithms}
  Chow, Y., Ghavamzadeh, M.: Algorithms for {CVaR} optimization in {MDPs}. Advances in neural information processing systems  \textbf{27} (2014)
  
  \bibitem{chow2015risk}
  Chow, Y., Tamar, A., Mannor, S., Pavone, M.: Risk-sensitive and robust decision-making: a cvar optimization approach. Advances in neural information processing systems  \textbf{28} (2015)
  
  \bibitem{CT18}
  Cubuktepe, M., Topcu, U.: Verification of {Markov} decision processes with risk-sensitive measures. In: Proc. Annual American Control Conference (ACC'18). pp. 2371--2377. IEEE (2018)
  
  \bibitem{DJKV17}
  Dehnert, C., Junges, S., Katoen, J.P., Volk, M.: A storm is coming: A modern probabilistic model checker. In: Proc. 29th International Conference on Computer Aided Verification (CAV'17) (2017)
  
  \bibitem{dpmc-nfm}
  Elsayed-Aly, I., Parker, D., Feng, L.: Distributional probabilistic model checking. In: Proc. 16th NASA Formal Methods Symposium (NFM'24). LNCS, Springer (2024)
  
  \bibitem{feng2016synthesis}
  Feng, L., Wiltsche, C., Humphrey, L., Topcu, U.: Synthesis of human-in-the-loop control protocols for autonomous systems. IEEE Transactions on Automation Science and Engineering  \textbf{13}(2),  450--462 (2016)
  
  \bibitem{filar1995percentile}
  Filar, J.A., Krass, D., Ross, K.W.: Percentile performance criteria for limiting average {Markov} decision processes. IEEE Transactions on Automatic Control  \textbf{40}(1),  2--10 (1995)
  
  \bibitem{HM14}
  Haddad, S., Monmege, B.: Reachability in {MDP}s: Refining convergence of value iteration. In: Proc. 8th International Workshop on Reachability Problems (RP'14). LNCS, vol.~8762, pp. 125--137. Springer (2014)
  
  \bibitem{HJKQ20}
  Hartmanns, A., Junges, S., Katoen, J.P., Quatmann, T.: Multi-cost bounded tradeoff analysis in {MDP}. Journal of Automated Reasoning  \textbf{64}(7),  1483--1522 (2020)
  
  \bibitem{JRSS18}
  Jha, S., Raman, V., Sadigh, D., Seshia, S.A.: Safe autonomy under perception uncertainty using chance-constrained temporal logic. Journal of Automated Reasoning  \textbf{60}(1),  43--62 (2018)
  
  \bibitem{klein2018advances}
  Klein, J., Baier, C., Chrszon, P., Daum, M., Dubslaff, C., Kl{\"u}ppelholz, S., M{\"a}rcker, S., M{\"u}ller, D.: Advances in probabilistic model checking with {PRISM}: variable reordering, quantiles and weak deterministic {B{\"u}chi} automata. International Journal on Software Tools for Technology Transfer  \textbf{20}(2),  179--194 (2018)
  
  \bibitem{kressgazit}
  Kress-Gazit, H., Fainekos, G.E., Pappas, G.J.: Temporal logic-based reactive mission and motion planning. IEEE Transactions on Robotics  \textbf{25}(6),  1370--1381 (2009)
  
  \bibitem{kvretinsky2018conditional}
  K{\v{r}}et{\'\i}nsk{\`y}, J., Meggendorfer, T.: Conditional value-at-risk for reachability and mean payoff in {Markov} decision processes. In: Proceedings of the 33rd Annual ACM/IEEE Symposium on Logic in Computer Science. pp. 609--618 (2018)
  
  \bibitem{KV01}
  Kupferman, O., Vardi, M.Y.: Model checking of safety properties. Formal Methods in System Design  \textbf{19}(3),  291–314 (2001)
  
  \bibitem{KNP11}
  Kwiatkowska, M., Norman, G., Parker, D.: {PRISM} 4.0: Verification of probabilistic real-time systems. In: Gopalakrishnan, G., Qadeer, S. (eds.) Proc. 23rd International Conference on Computer Aided Verification (CAV'11). LNCS, vol.~6806, pp. 585--591. Springer (2011)
  
  \bibitem{KNP12b}
  Kwiatkowska, M., Norman, G., Parker, D.: The {PRISM} benchmark suite. In: Proc. 9th International Conference on Quantitative Evaluation of SysTems (QEST'12). pp. 203--204. IEEE CS Press (2012)
  
  \bibitem{lyle2019comparative}
  Lyle, C., Bellemare, M.G., Castro, P.S.: A comparative analysis of expected and distributional reinforcement learning. In: Proceedings of the AAAI Conference on Artificial Intelligence. vol.~33, pp. 4504--4511 (2019)
  
  \bibitem{majumdar2020should}
  Majumdar, A., Pavone, M.: How should a robot assess risk? {Towards} an axiomatic theory of risk in robotics. In: Robotics Research: The 18th International Symposium ISRR. pp. 75--84. Springer (2020)
  
  \bibitem{meggendorfer2022risk}
  Meggendorfer, T.: Risk-aware stochastic shortest path. In: Proceedings of the AAAI Conference on Artificial Intelligence. vol.~36, pp. 9858--9867 (2022)
  
  \bibitem{Pnu81}
  Pnueli, A.: The temporal semantics of concurrent programs. Theoretical Computer Science  \textbf{13},  45--60 (1981)
  
  \bibitem{randour2017percentile}
  Randour, M., Raskin, J.F., Sankur, O.: Percentile queries in multi-dimensional {Markov} decision processes. Formal methods in system design  \textbf{50},  207--248 (2017)
  
  \bibitem{rigter2022planning}
  Rigter, M., Duckworth, P., Lacerda, B., Hawes, N.: Planning for risk-aversion and expected value in {MDPs}. In: Proceedings of the International Conference on Automated Planning and Scheduling. vol.~32, pp. 307--315 (2022)
  
  \bibitem{rockafellar2002conditional}
  Rockafellar, R.T., Uryasev, S.: Conditional value-at-risk for general loss distributions. Journal of banking \& finance  \textbf{26}(7),  1443--1471 (2002)
  
  \bibitem{sobel1982variance}
  Sobel, M.J.: The variance of discounted {Markov} decision processes. Journal of Applied Probability  \textbf{19}(4),  794--802 (1982)
  
  \bibitem{UB13}
  Ummels, M., Baier, C.: Computing quantiles in {Markov} reward models. In: Proc. 16th International Conference on Foundations of Software Science and Computation Structures (FOSSACS'13). LNCS, vol.~7794, pp. 353--368. Springer (2013)
  
  \end{thebibliography}

}{%

\newpage
\appendix
\setcounter{lemma}{1}

\section*{Appendix}

We provide a proof for Lemma~\ref{lem:slack_variable}
relating to the correctness of our method in Section~\ref{sec:cvar}.
Let $X_{\mdp,\policy}^{r,\futureop\,T}$ be the random variable
for accumulation of reward $r$ until reaching states $T$
under policy $\pi$ of an MDP $\mdp$.

We first restate a closely related theorem from~\cite{bauerle2011markov}.

\begin{theorem}
~\cite[Theorem 4.5, adapted]{bauerle2011markov} \label{thm:convergence}
Assuming a continuous slack variable $b \in \Rset$,
there exists a solution $\binit^*$ when minimizing, over values of $b$, the inner formula $\exp^\pi([X_{\mdp,\policy}^{r,\futureop\,T} - b ]^+)$ of Equation \ref{eqn:cvar_duel},
for which the optimal policy for 
$\CVaR{\alpha}(X_{\mdp^b,\pi}^{r,\futureop\,T})$
from initial state $\saug{\sinit}{\binit^*}$ in the augmented MDP $\mdp^b$ solves the following optimization problem for a fixed $\alpha \in [0,1]$ in the original MDP $\mdp$:
$$\inf _{\pi \in \Pi} \CVaR{\alpha}(X_{\mdp,\pi}^{r,\futureop\,T})
$$ 
\end{theorem}

Theorem \ref{thm:convergence} guarantees that risk-sensitive DVI using the inner formula for the augmented MDP $\mdp^b$ leads to the optimal policy. 
Moreover, the optimal policy for the augmented model $\mdp^b$ is also optimal for the original MDP $\mdp$. 
However, this theorem assumes a continuous slack variable. 
In the following, we prove that, as the number of atoms used for $|B|$ increases (i.e., stride $\varsigma_n$ decreases), the optimal policy for a finite set $B$ becomes closer to the optimal policy with a continuous slack variable in terms of CVaR values.

\begin{lemma}
\label{lem:slack_variable_full}
    Let $\pi_1$ denote the optimal policy for minimizing $\CVaR{\alpha}(X_{\mdp,\policy}^{r,\futureop\,T})$, which is obtained with a continuous slack variable.
    Let $\pi_2$ denote the optimal policy returned by \agref{ag:cvar} where $B$ is a finite set of $n$ evenly-spaced atoms with stride $\varsigma_n$. It holds that
    $\CVaR{\alpha}(X_{\mdp,\pi_2}^{r,\futureop\,T}) - \CVaR{\alpha}(X_{\mdp,\pi_1}^{r,\futureop\,T}) = \cO(\varsigma_n)$. As $\varsigma_n$ tends to 0 (i.e., $|B|$ increases), $\pi_2$ converges to the $\CVaR{}$ optimal policy.
\qed
\end{lemma}

\begin{proof}
Recall that, when building the product MDP in \agref{ag:cvar}, we determine the slack variable value $b'$ for a successor state by rounding down the value of $b-r(s,a)$ to the nearest smaller atom in $B$. 
More precisely, 
\begin{eqnarray*}
b'=  \Vmin +\varsigma_n \cdot \left\lfloor \frac{ \max (\Vmin, b-r) - \Vmin}{\varsigma_n } \right\rfloor 
\end{eqnarray*}

This is the main source of approximation errors introduced by the discretization of the slack variable.  
Following the dual representation of $\CVaR{}$ given in \lemref{lem:cvar-duel}, each slack variable value update would introduce the error of

\begin{eqnarray*}
& & \CVaR{\alpha}(X_{\mdp,\pi_2}^{r,\futureop\,T}) - \CVaR{\alpha}(X_{\mdp,\pi_1}^{r,\futureop\,T}) \\
&=& \big( b' - \max (\Vmin, b-r) \big) \\
&+& \frac{1}{1-\alpha} \left(\exp (\left[ X - b'\right]^+) - \exp (\left[ X -\max (\Vmin, b-r) \right]^+) \right)
\end{eqnarray*}

Let $\beta := \max (\Vmin, b-r)$. The first term yields, 
\begin{eqnarray*}
b' - \beta 
&=&  \Vmin + \varsigma_n \cdot \left\lfloor \frac{\beta - \Vmin}{\varsigma_n } \right\rfloor - \beta \\
&=& \varsigma_n \cdot \left( \left\lfloor \frac{\beta- \Vmin}{\varsigma_n } \right\rfloor - \frac{\beta -\Vmin}{\varsigma_n} \right) \\
&\in& (-\varsigma_n , 0]
\end{eqnarray*}

Given $x_1, x_2, x_3 \in \Rset$, we have 
$\left[x_1-x_2\right]^+ - \left[x_1-x_3 \right]^+  \leq \left[x_3-x_2\right]^+$
based on the triangle inequality.
The second term yields

\begin{eqnarray*}
& & \frac{1}{1-\alpha} \left(\exp (\left[ X - b'\right]^+) - \exp (\left[ X - \beta \right]^+) \right) \\
&=& \frac{1}{1-\alpha} \exp \left( \left[ X - b'\right]^+ - \left[ X - \beta \right]^+  \right) \\
&\le& \frac{1}{1-\alpha} \exp \left( \left[ \beta - b' \right]^+ \right) \\
&=& \frac{1}{1-\alpha} \exp \left( \left[ \beta - \Vmin -\varsigma_n \cdot \left\lfloor \frac{\beta -\Vmin}{\varsigma_n } \right\rfloor \right]^+ \right) \\
&=& \frac{\varsigma_n}{1-\alpha} \exp \left( \left[ \frac{\beta-\Vmin}{\varsigma_n} - \left\lfloor \frac{\beta -\Vmin}{\varsigma_n } \right\rfloor  \right]^+ \right) \\
&\in&[0, \frac{\varsigma_n}{1-\alpha})
\end{eqnarray*}

Thus, it holds that $\CVaR{\alpha}(X_{\mdp,\pi_2}^{r,\futureop\,T}) - \CVaR{\alpha}(X_{\mdp,\pi_1}^{r,\futureop\,T}) = \cO(\varsigma_n)$.
As $\varsigma_n$ tends to 0 (i.e., $|B|$ increases), $\pi_2$ converges to the $\CVaR{}$ optimal policy.
\qed
\end{proof}

Finally, we comment on the use of (categorical or quantile) distributional representations
and projections for implementing the above. 
Note that \cite[Proposition 5.28]{bookdr2022} proves that using the distributional Bellman update with a distributional representation/projection combination (categorical or quantile) is guaranteed to converge based on the stride and the number of iterations.

}

\end{document}